\documentclass[iop]{emulateapj}

\usepackage{amsmath}
\usepackage{hyperref}  
\usepackage{breakurl}  
\newcommand\kms{\ifmmode{\rm km\thinspace s^{-1}}\else km\thinspace s$^{-1}$\fi}

\newcommand{\hd}{HD\,98800}
\newcommand{\hda}{HD\,98800\,A}
\newcommand{\hdb}{HD\,98800\,B}

\shortauthors{Torres et al.}
\shorttitle{\hd}

\begin{document}
\submitted{Accepted for publication in The Astrophysical Journal}

\title{Refining the Architecture of the quadruple pre-main-sequence system \hd: Improved Orbits, Dynamical Masses, and a Transiting Circumbinary Disk}

\author{
Guillermo Torres\altaffilmark{1},         
John D.\ Monnier\altaffilmark{2},         
Gail H.\ Schaefer\altaffilmark{3},        
Andrew F.\ Boden\altaffilmark{4},         
Matthew J.\ Holman\altaffilmark{1},\\     
Stefan Kraus\altaffilmark{5},             
Noura Ibrahim\altaffilmark{2},            
James W.\ Davidson, Jr.\altaffilmark{6},  
Elliott P.\ Horch\altaffilmark{7,8},      
Lisa Prato\altaffilmark{9,10},\\          
David W.\ Latham\altaffilmark{1},         
Robert P.\ Stefanik\altaffilmark{1},      
Narsireddy Anugu\altaffilmark{11}, and    
Claire L.\ Davies\altaffilmark{5}         
}

\altaffiltext{1}{Center for Astrophysics $\vert$ Harvard \& Smithsonian, 60 Garden St., Cambridge, MA 02138, USA; gtorres@sao.si.edu}
\altaffiltext{2}{Astronomy Department, University of Michigan, Ann Arbor, MI 48109, USA}
\altaffiltext{3}{The CHARA Array of Georgia State University, Mount Wilson Observatory, Mount Wilson, CA 91023, USA}
\altaffiltext{4}{California Institute of Technology, Mail Code 11-17, 1200 East California Blvd., Pasadena, CA 91125, USA}
\altaffiltext{5}{Astrophysics Group, Department of Physics \& Astronomy, University of Exeter, Exeter, EX4 4QL, UK}
\altaffiltext{6}{Department of Astronomy, University of Virginia, 530 McCormick Road, Charlottesville, VA 22904, USA}
\altaffiltext{7}{Department of Physics, Southern Connecticut State University, 501 Crescent Street, New Haven, CT 06515, USA}
\altaffiltext{8}{Adjunct Astronomer, Lowell Observatory.}
\altaffiltext{9}{Physics \& Astronomy Department, Rice University, 6100 Main St., Houston, TX 77005, USA}
\altaffiltext{10}{Space Science Institute, 4765 Walnut St, Suite B, Boulder, CO 80301, USA}
\altaffiltext{11}{The CHARA Array of Georgia State Univerity, Mount Wilson Observatory, Mount Wilson, CA 91203, USA}

\begin{abstract}
We present new spectroscopic and interferometric observations of the
pre-main-sequence hierarchical quadruple system \hd, a member of
the nearby TW\,Hya association. Our radial velocity monitoring extends
the total coverage to over 36 years, sampling the periastron passage
in the $\sim$240-yr outer orbit. By combining these velocities with new and
archival interferometric observations of both inner binaries (A and
B), we significantly refine the system orbits and constrain the
dynamical masses of all four stars to $\leq$ 2.6\%. The resulting
stellar properties (masses, effective temperatures, and absolute
magnitudes) are found to agree well with current stellar evolution
models.  Notably, binary B hosts a known circumbinary disk in a polar
orientation. Our interferometric imaging, as well as ground-based
photometry by others, capture binary A currently passing directly
behind the disk, a striking alignment given the outer orbit’s long
period. We use the photometry to infer the optical extinction produced
by the disk material. The eclipse is expected to last about a decade,
and offers an unprecedented opportunity to probe the structure and
composition of the potentially planet-forming disk. Our refined orbits
and component masses provide a robust dynamical framework to support
ongoing and future studies of this unique event.
\end{abstract}


\section{Introduction}
\label{sec:introduction}

Pre-main-sequence (PMS) stars are known to often appear in
gravitationally bound pairs, or in systems of higher multiplicity, more
so than older main-sequence objects \citep[e.g.,][]{Duchene:2013}. A fascinating example is
the hierarchical quadruple PMS system \hd\ (TV\,Crt), which is a member of the
nearby $\sim$10 Myr old TW Hya association. This is a loose grouping of
several dozen young stars covering a wide area on the sky
\citep{Luhman:2023}, within which \hd\ has received the designation TWA-4
{\citep{Webb:1999}}.
Visual binary observers during the last century found it to be double,
with a north-south separation of about 0\farcs8 at the time. Much
later, each visual component was in turn discovered to be a
spectroscopic binary \citep{Torres:1995}: \hda, the southern
component, is single-lined\footnote{We take this opportunity to
correct a misstatement in Table~15 of \cite{Torres:2003}, where
\hda\ was listed as being double-lined. So far as we are aware, it has only been seen as
single-lined.} and has
an orbital period of 264 days, while \hdb\ is double-lined, with a
period of 314 days. Both orbits are considerably eccentric.

In the 1980s it was found that the object displays an unusually
large infrared excess \citep{Walker:1988}, which was later shown to be
caused by a circumbinary disk around the B binary \citep{Gehrz:1999,
Koerner:2000, Prato:2001}.  In the quarter of a century since, \hd\
has attracted considerable attention in an effort to better understand
the properties of the disk \citep{Akeson:2007, Furlan:2007,
Verrier:2008, Andrews:2010, Ribas:2018, Kennedy:2019, Ronco:2021,
Faruqi:2025, Rabago:2025}.
It is now known, from the imaging observations of \cite{Kennedy:2019} at 1.3\,mm with the Atacama Large Millimeter/submillimeter Array (ALMA), that the disk is strongly misaligned relative to the orbit of the binary, residing in a polar configuration. This is the first recorded example of a protoplanetary disk with this orientation. According to these authors, the dust component of the disk is truncated at its outer edge at about 4.6~au, presumably by the perturbative action of \hda, and has a central cavity with an inner disk radius of $\sim$2.5\,au, carved out by the inner binary. The gas distribution, detected from the CO $J =2$--1 rotational transition, is more extended.

Efforts have also been undertaken to improve our knowledge of the inner and
outer orbits. The components of B were first spatially resolved by
\cite{Boden:2005} with the Keck Interferometer (KI), and those of A
were later resolved by \cite{Zuniga:2021} using the Very Large
Telescope Interferometer (VLTI). Both inner orbits have semimajor axes
of about 20~mas. The elements of the outer orbit, which are less well constrained, have also been revisited by \cite{Tokovinin:1999}, \cite{Tokovinin:2014},
\cite{Kennedy:2019}, and \cite{Zuniga:2021, Zuniga:2026}, with the latter authors
estimating the period to be 212--237\,yr in their most recent work.

A remarkable feature of the \hd\ system is that the outer orbit
is nearly edge-on, and the A and B components, currently separated by about 0\farcs13, are closing in on each other, with the B
binary located in front of the other. It was realized a few years ago
\citep[e.g.,][]{Kennedy:2019} that this configuration would lead to the A binary passing directly behind the circumbinary disk around B, offering a unique
opportunity to study the structure and composition of the disk. The event has already started and will last several years. \cite{Zuniga:2021} and \cite{Faruqi:2025}
have simulated the light curve
expected throughout the event. The details depend on both the 
properties of the disk and the elements of the outer orbit.

The component stars themselves have also been characterized to the extent possible.
Dynamical mass estimates for the B components that are good to about 10\%
were previously reported by \cite{Boden:2005}, and were updated and
improved to the level of 2--3\% by \cite{Zuniga:2026}. Those of the A components are less well known (4--6\%), limited by the single-lined nature of the
spectroscopic orbit.

The spectroscopic monitoring of \hd\ at the Center for Astrophysics
(CfA) by \cite{Torres:1995}, which led to the discovery of the binary
nature of each of the visual components, has been continued and now
spans more than 36\,yr. The data provide full coverage of the most recent 2021 periastron
passage in the outer orbit. In addition, we have obtained new long-baseline
interferometric observations of the system, and taken advantage of several archival interferometric
observations that are available as well. A number of speckle measurements
of the visual pair have also been collected in recent years. Together,
these extended datasets have motivated us to revisit the
orbital solutions, and the determination of the component masses, with an eye toward
providing a much improved dynamical picture of the system for the
study of the ongoing occultation of \hda\ by the circumbinary disk around
the other binary.

The plan for the paper is as follows. Section~\ref{sec:observations} reports the new spectroscopic and astrometric observations, as well as other published radial-velocity measurements and archival interferometric observations that we have used in our analysis. In Section~\ref{sec:orbit}, we describe and present the results of our joint orbital solution of the three orbits, in which we introduce various functional relationships among the elements that allow us to reduce the number of adjustable parameters for this particular case. In Section~\ref{sec:discussion} we discuss {our findings}, including updated estimates of the relative inclination angles among the three orbits, and the dynamical evolution of the system. {We also quantify temporal} changes in the inner orbital elements, both in the vicinity of the 2021 periastron passage as well as evolution over the longer term. This section {includes a comparison of} various properties of the components against current models of stellar evolution for PMS stars. We also report observational evidence that the {occultation} of \hda\ is in progress, and we describe our use of ground-based photometry to infer the optical extinction produced by the disk so far. A brief discussion follows about the possibility of an eclipse of the secondary of \hda\ by the primary of \hdb\ in late 2028. We conclude with final remarks in Section~\ref{sec:conclusions}.
 

\section{Observations and Reductions}
\label{sec:observations}

\subsection{CfA Spectroscopy}
\label{sec:spectroscopy}

The observations originally reported by \cite{Torres:1995} that 
revealed \hda\ and B to be spectroscopic binaries were gathered at
the CfA over the interval 1989.9--1994.4. They were obtained at three
different telescopes, as described in that work, using three copies of the
Digital Speedometer \citep[DS;][]{Latham:1985, Latham:1992}. These
echelle instruments were equipped with intensified photon-counting
Reticon detectors, and recorded a single order about 45\,\AA\ wide
centered on the \ion{Mg}{1}\,b triplet near 5187~\AA, at a resolving
power of $R \approx 35,000$. The \hd\ spectra from \cite{Torres:1995}
typically had low signal-to-noise ratios (SNRs) between 6 and 23 per
resolution element of 8.5~\kms, with a median of 12. At the time, only
152 of the 177 spectra available were considered to yield acceptable
radial-velocity (RV) measurements.

After a five year pause, monitoring was continued for another six
years with the same instrumentation (1999.1--2006.9), and a further 59
spectra were collected with longer exposure times. The SNRs were
correspondingly higher, with a median of about 32 per resolution
element (range = 17--48). As with the previous spectra, the reductions
followed the procedures described by \cite{Latham:1985, Latham:1992}.

Starting in 2018, we obtained another set of spectra using the
Tillinghast Reflector Echelle Spectrograph
\citep[TRES;][]{Furesz:2008, Szentgyorgyi:2007} on the 1.5\,m telescope
at the Fred L.\ Whipple Observatory (Mount Hopkins, AZ, USA). This is
a modern, bench-mounted, fiber-fed instrument with a CCD detector,
delivering a resolving power of $R \approx 44,000$ in 51 orders, over
the wavelength range 3800--9100~\AA. The SNRs for \hd\ are
considerably higher with this instrument, with a median of 67 per resolution element of
6.8~\kms\ (range = 32--139).  The reduction of the 41 TRES spectra,
gathered between 2018.0 and 2026.1, was performed with a dedicated
pipeline \citep[see][]{Buchhave:2012}.

The velocity zero-point of each instrument was monitored every night
with observations of the twilight sky (in the case of the DS) or IAU
standards (TRES), and small adjustments were made to place all
measurements on the same native CfA reference frame. Observations of
minor planets in the Solar System were then used to transfer the
velocities described below onto the IAU system.

All of our spectra of \hd\ are of the combined light of the two visual
components, as the angular separation is too small to split them at
the telescope under typical observing conditions. Three sets of lines
are present, corresponding to the primary of binary A and both components
of binary B.
RVs were measured in the same way as described by
\cite{Torres:1995}, using the three-dimensional cross-correlation
algorithm {\tt TRICOR} \citep{Zucker:1995}, which is designed to
overcome biases due to line blending in composite spectra.
For TRES, we chose to use
the order centered on the \ion{Mg}{1}\,b triplet (100~\AA\ wide), for
consistency with the DS, and also because experience has shown that
this region of the spectrum contains most of the velocity information.
The templates (one for each of the visible components) were selected
from a pre-computed library of synthetic spectra based on Kurucz model
atmospheres \cite[see][]{Nordstrom:1994, Latham:2002}, and a line list
tuned manually by Jon Morse to improve the agreement with real
stars. After considerable experimentation, we settled on templates
with temperatures of 4500~K for the primaries of both spectroscopic
binaries (stars Aa and Ba), and 4000~K for star Bb. We adopted surface
gravities of $\log g = 4.0$ for the three stars, along with solar
metallicity. The line broadenings are small.  Following the determinations
of \cite{Soderblom:1996}, and within the limitations of our template
grid, we adopted $v \sin i$ values of 4, 2, and 0~\kms\ for Aa, Ba,
and Bb, respectively.

To take advantage of improvements in our analysis software over the
years, we rederived the RVs from the spectra originally reported by
\cite{Torres:1995} in the same way as for the more recent DS and TRES
spectra. Therefore, {those published 1995 velocities are superseded by the ones derived here.} Additionally, examination of the residuals from
our global orbital solution, described later in
Section~\ref{sec:orbit}, showed systematic RV deviations for the fainter
of the three stars (Bb) from the weakest of the DS spectra.
Consequently, for that star we chose to use only DS spectra with SNRs
higher than 10. The velocities we derived, along with their internal
errors, are listed in Table~\ref{tab:rvs}. Altogether, they provide
a baseline of more than 36\,yr.  The flux ratios between the components,
corresponding to the mean wavelength of our observations (5187~\AA), are
as follows: $\ell_{\rm Ba}/\ell_{\rm Aa} = 0.33 \pm 0.04$ and
$\ell_{\rm Bb}/\ell_{\rm Aa} = 0.12 \pm 0.02$ from the DS spectra (45\,\AA\ bandwidth), and $\ell_{\rm Ba}/\ell_{\rm Aa} = 0.53 \pm 0.08$ and $\ell_{\rm
Bb}/\ell_{\rm Aa} = 0.15 \pm 0.03$ from TRES (100\,\AA\ bandwidth).

\setlength{\tabcolsep}{6pt}
\begin{deluxetable*}{lcccccccccc}
\tablewidth{0pc}
\tablecaption{CfA RV Measurements for \hd\ \label{tab:rvs}}
\tablehead{
\colhead{BJD} &
\colhead{RV$_{\rm Aa}$} &
\colhead{RV$_{\rm Ba}$} &
\colhead{RV$_{\rm Bb}$} &
\colhead{$\sigma_{\rm Aa}$} &
\colhead{$\sigma_{\rm Ba}$} &
\colhead{$\sigma_{\rm Bb}$} &
\colhead{$\tau_{\rm A}$} &
\colhead{$\tau_{\rm B}$} &
\colhead{Instr.}
\\
\colhead{(2,400,000+)} &
\colhead{(\kms)} &
\colhead{(\kms)} &
\colhead{(\kms)} &
\colhead{(\kms)} &
\colhead{(\kms)} &
\colhead{(\kms)} &
\colhead{(day)} &
\colhead{(day)} &
\colhead{}
}
\startdata
 47849.9288 & 15.32 & 2.12 & \nodata    & 1.43 & 3.11 & \nodata & 0.0958 & $-0.0848$ & 1 \\
 47893.9437 & 16.52 & 3.35 & \nodata	& 1.33 & 2.91 & \nodata & 0.0952 & $-0.0843$ & 1 \\
 47900.9284 & 18.27 & 6.67 & \nodata	& 1.48 & 3.23 & \nodata & 0.0951 & $-0.0842$ & 1 \\
 47923.8619 & 20.46 & 7.72 & \nodata	& 1.38 & 3.01 & \nodata & 0.0948 & $-0.0840$ & 1 \\
 47933.7875 & 20.18 & 4.14 & \nodata	& 1.19 & 2.60 & \nodata & 0.0946 & $-0.0838$ & 1
\enddata

\tablecomments{$\tau_{\rm A}$ and $\tau_{\rm B}$ represent the light travel
time corrections added to the times of observation, to transfer them to the
frame of the center of mass of the quadruple system. The instrument code
indicated in the last column is 1 for the Digital Speedometer, and 2 for TRES.
(This table is available in its entirety in machine-readable form.)}

\end{deluxetable*}
\setlength{\tabcolsep}{6pt}

\subsubsection{Other published RVs}
\label{sec:otherRVs}

Several other RV datasets for \hd\ have been reported in the
literature \citep[e.g.,][]{Zuniga:2021, Zuniga:2026}, but they are generally small (from a handful to a dozen
measurements), and in some cases the spectra do not resolve the three
components. These measurements typically contribute little to our
analysis, particularly when they overlap in time with ours. The one
exception we have allowed is the set of velocities by
\cite{Merle:2024}, consisting of two dozen measurements for star Aa
and a smaller number for Ba and Bb, all obtained with the same
instrumentation (the HERMES spectrograph on the 1.2\,m Mercator
telescope at La Palma, Spain). These velocities are all ostensibly on the same velocity
reference frame, close to the IAU system. They happen to
fill an 11\,yr gap in our own RV coverage, just before
periastron passage in the outer orbit. We adopted them as published.

\subsection{Long-baseline Interferometry}
\label{sec:longbaseline}

Near-infrared, long-baseline interferometric observations of \hd\ have
been gathered over more than 20~yr. Some are published, others we have
extracted from data archives for use in our analysis, and some are
new.

\subsubsection{Keck Interferometer}
\label{sec:KI}

The earliest interferometric observations of this system were published by \cite{Boden:2005}, and resolved the B component for the first time.  They were obtained in the $K$ band with the KI on five nights between 2003 and 2005. The data published are the squared visibilities (fringe contrast). From their set of 34 calibrated visibility measurements, the authors reported an astrometric orbit that also incorporated other astrometric measurements, as well as the RVs for \hdb\ from \cite{Torres:1995}. In the course of a preliminary orbital analysis for \hd, we encountered unexpected difficulties in using these observations, which appeared to be in tension with the larger set of RVs we now have. Our own independent orbital solution based on the same data sets used by \cite{Boden:2005} confirmed the accuracy of their elements, i.e., that their published orbit model fits the published KI measurements well. Surprisingly, we found that new interferometric measurements from CHARA and the VLTI, reported below, deviated significantly from that orbit by several milli-arcseconds (mas) in a systematic fashion. 

KI observations of \hdb\ were continued after the 2005 publication, and a total of seven additional nights of data were gathered through 2010. Beginning in 2004, \hda\ was observed with the KI as well on ten separate nights, many in common with \hdb, and the Aa and Ab components were clearly resolved. The observing protocols were similar to those applied to B \citep[see][]{Boden:2005}. As indicated there, the Keck telescopes resolve the \hd\,AB system (separation $\sim0\farcs75$ at the time), and the KI beam combiner was fed by a single-mode fiber such that, when observing one of the visual components, no light from the other fell on the tracking camera. Both \hda\ and \hdb\ were observed several times a night together with calibration sources. For B, the calibrators were HD\,97590 (\ion{A6}{5}) and HD\,99256 (\ion{A9}{5}). For A, the calibration source was HD\,97590.\footnote{In reviewing the suitability of these objects as calibrators based on the latest information available, we found that the Gaia DR3 multi-source classifier (MSC) declared HD\,97590 to be a non-single star with near 100\% probability ({\tt classprob\_dsc\_combmod\_star} flag). This assessment relied largely on the BP/RP spectrum from the mission. While this might seem to disqualify it as a calibrator, and raise the concern that it could be playing a role in the discrepancies noted above for \hdb, the ``binarity'' claim is not to be taken literally, as explained, e.g., by \cite{Delchambre:2023}. This metric was not designed to be used to distinguish binary from single stars, and it performs poorly in that regard. We have found no other evidence of binarity in the literature, or in our own assessments dating back to 2000, and in fact, the newly calibrated visibilities for binary B change very little when we exclude HD\,97590, and use only the other calibrator.} The additional observations for B, and those for A, have not been previously published, so we report them here. For consistency, we have re-reduced all of the KI data adopting updated uniform-disk angular diameters for the calibrators of $\phi = 0.170 \pm 0.005$\,mas (HD\,97590) and $\phi = 0.199 \pm 0.007$\,mas (HD\,99256), based on new modeling of the spectral energy distributions using considerably more radiometry than was available in 2005. We present the recalibrated visibility measurements in Table~\ref{tab:keckA} and Table~\ref{tab:keckB} (104 for A, 92 for B), with other ancillary information. The measurements for B supersede those by \cite{Boden:2005}.

The difficulties found with the original visibilities for B persisted to some extent with the new calibration. The measurements for A showed much less of a problem, although there still appeared to be some excess scatter. One of the nights for \hdb\ seemed more discrepant than others (MJD 53101). These signs are suggestive of systematic uncertainties not captured by the internal errors, which may have caused subtle biases in the \cite{Boden:2005} orbit. We discuss this in more detail in Section~\ref{sec:orbit}.

\setlength{\tabcolsep}{5pt}
\begin{deluxetable}{lccccc}
\tablewidth{\columnwidth}
\tablecaption{Recalibrated KI Visibility Measurements for \hda\ \label{tab:keckA}}
\tablehead{
\colhead{MJD} &
\colhead{$\lambda$} &
\colhead{$V^2$} &
\colhead{$\sigma_{V^2}$} &
\colhead{$U$} &
\colhead{$V$}
\\
\colhead{} &
\colhead{($\mu$m)} &
\colhead{} &
\colhead{} &
\colhead{(m)} &
\colhead{(m)}
}
\startdata
53101.37038 & 2.147 & 0.448 & 0.096 & 50.755468 & 47.152672 \\
53101.37256 & 2.147 & 0.627 & 0.067 & 50.408248 & 46.860976 \\
53101.37478 & 2.147 & 0.636 & 0.071 & 50.044794 & 46.565956 \\
53101.37602 & 2.147 & 0.608 & 0.048 & 49.838581 & 46.402948 \\
53101.39270 & 2.147 & 0.556 & 0.051 & 46.761954 & 44.270912 
\enddata

\tablecomments{The $V^2$ uncertainties, $\sigma_{V^2}$, are the values generated by the calibration pipeline, and $U$ and $V$ represent the interferometric baselines. (This table is available in its entirety in machine-readable form.)}

\end{deluxetable}
\setlength{\tabcolsep}{6pt}

\setlength{\tabcolsep}{5pt}
\begin{deluxetable}{lccccc}
\tablewidth{\columnwidth}
\tablecaption{Recalibrated KI Visibility Measurements for \hdb\ \label{tab:keckB}}
\tablehead{
\colhead{MJD} &
\colhead{$\lambda$} &
\colhead{$V^2$} &
\colhead{$\sigma_{V^2}$} &
\colhead{$U$} &
\colhead{$V$}
\\
\colhead{} &
\colhead{($\mu$m)} &
\colhead{} &
\colhead{} &
\colhead{(m)} &
\colhead{(m)}
}
\startdata
52747.34508 & 2.147 & 0.728 & 0.040 & 49.876150 & 46.438393 \\
52747.34625 & 2.147 & 0.724 & 0.056 & 49.677133 & 46.283438 \\
52747.36253 & 2.147 & 0.686 & 0.043 & 46.651992 & 44.209811 \\
52747.36370 & 2.147 & 0.652 & 0.039 & 46.415618 & 44.066156 \\
52747.37528 & 2.147 & 0.632 & 0.046 & 43.938414 & 42.682951 
\enddata

\tablecomments{The $V^2$ uncertainties, $\sigma_{V^2}$, are the values generated by the calibration pipeline, and $U$ and $V$ represent the interferometric baselines. (This table is available in its entirety in machine-readable form.)}

\end{deluxetable}
\setlength{\tabcolsep}{6pt}

\subsubsection{CHARA}
\label{sec:chara}

Table~\ref{tab:chara_log} lists the UT dates of the archival and new observations of \hd\ collected at the CHARA Array \citep{tenBrummelaar:2005} using the MIRC-X beam combiner in the $H$ band \citep{Anugu:2020} and the MYSTIC combiner in the $K$ band \citep{Setterholm:2023}. The data were collected by combining the light from four or five telescopes together simultaneously (W1, W2, S1, S2, E2). The E1 telescope was not in delay range because of the target's low declination. Most observations were taken in the low spectral resolving power mode ($R \sim 50$); a few were taken with higher spectral resolving power ($R \sim 100$) which provides a wider interferometric field of view (see Section~\ref{sec:charavlti}). Observations of unresolved calibrator stars were interspersed before and/or after the observations of \hd. The instrument, spectral mode, telescope configuration, and calibrator stars are listed in Table~\ref{tab:chara_log}.

The angular separation between \hda\ and B varied from 0$\farcs$34 to 0$\farcs$12 over CHARA observations from 2021 to 2026. Therefore, both wide components fell within the diffraction limited field of view of a single 1\,m CHARA telescope (0\farcs4 at $H$ and 0\farcs5 at $K$), and the telescopes were presumably centered on the brighter A component. However, light from all four components was recorded on the detector.

The MIRC-X and MYSTIC data were reduced with the standard pipeline\footnote{\url{https://gitlab.chara.gsu.edu/lebouquj/mircx_pipeline}} (Version 1.5.0). The data were reduced using an integration time of 30 sec to use the deep cleaning option for removing outliers with the IDL calibration script written by J.\ D.\ Monnier. The data were then averaged over 5 minute and 2.5 minute intervals, and also reduced with no averaging. We also used a longer integration time of 150 sec to carry out the calibration using the default python calibration script that is packaged with the pipeline. The results from the different calibration scripts {were consistent}. To avoid time smearing effects when fitting for the {inner} binary components, we selected the shortest averaging time interval that produced a signal-to-noise ratio suitable for measuring reliable binary positions. Table~\ref{tab:chara_cal} lists the angular diameters in $H$ and $K$ that we adopted for the calibrators, from the JMMC Stellar Diameter Catalog \citep{Chelli:2016,Bourges:2017}, along with their 1$\sigma$ uncertainties.

\setlength{\tabcolsep}{6pt}
\begin{deluxetable*}{llllll}
\tablewidth{\textwidth}
\tablecaption{Log of Observations at the CHARA Array \label{tab:chara_log}}
\tablehead{
\colhead{UT Date}   &
\colhead{Instrument} &
\colhead{$\lambda$} &
\colhead{Mode}     &
\colhead{Telescopes}      &
\colhead{Calibrator HD Numbers}
}
\startdata
2021 May 11 & MIRC-X     & $H$         & Prism50  & W2-W1-S2-S1-E2  & 100110, 138856, 153118, 145375, 156365 \\
2021 May 12 & MIRC-X     & $H$         & Prism50  & W2-W1-S2-S1-E2  & 100110, 135969, 145965, 139487, 151259 \\
2021 May 14 & MIRC-X     & $H$         & Prism50  & W2-W1-S2-E2     & 98219, 135969  \\
2021 May 15\tablenotemark{a} & MIRC-X     & $H$         & Prism50  & W2-W1-S2-E2     & 100110, 135969  \\
2023 Apr 27\tablenotemark{b} & MIRC-X     & $H$         & Prism50  & E2-W2-W1-S1-S2  & 98729, 103155, 153213 \\
2023 Apr 28\tablenotemark{b} & MIRC-X     & $H$         & Prism50  & E2-W2-W1-S1-S2  & 98729, 103155, 163843 \\
2023 May 01\tablenotemark{a,b} & MIRC-X     & $H$         & Prism50  & E2-W2-W1-S1     & 98729, 141162  \\
2024 May 15\tablenotemark{a} & MIRC-X     & $H$         & Prism102 & W2-W1-S2-S1-E2  & 91953, 98219, 107662 \\
2024 May 15\tablenotemark{a} & MYSTIC     & $K$         & Prism49  & W2-W1-S2-S1-E2  & 91953, 98219, 107662 \\
2024 May 28 & MIRC-X     & $H$         & Prism102 & W2-W1-S2-S1-E2  & 105395, 152557, 169580, 175120, 170347, 178834 \\
2024 May 28 & MYSTIC     & $K$         & Prism49  & W2-W1-S2-S1-E2  & 105395, 152557, 169580, 175120, 170347, 178834 \\
2025 May 03 & MIRC-X     & $H$         & Prism50  & W2-S2-S1-E2     & 103580 \\
2025 May 03 & MYSTIC     & $K$         & Prism49  & W2-S2-S1-E2     & 103580 \\
2026 Mar 25\tablenotemark{c} & MIRC-X     & $H$         & Prism50  & W2-W1-S2-S1-E2  & 94322, 103635  \\
2026 Mar 25\tablenotemark{c} & MYSTIC     & $K$         & Prism100 & W2-W1-S2-S1-E2  & 94322, 103635 
\enddata
\tablenotetext{a}{No reliable solution was found on UT 2021\,May\,15, 2023\,May\,01, and 2024\,May\,15 due to sparse coverage and/or low SNR.}
\tablenotetext{b}{On UT 2023\,Apr\,27, 2023\,Apr\,28, and 2023\,May\,01 the MYSTIC data were collected with $K$ Prism49 in the Gravity ABCD-4T mode. The data from those nights have not been reduced because the pipeline for this mode is not yet ready.}
\tablenotetext{c}{The data on UT 2026\,Mar\,25 consist of two 10-minute data sets.}
\end{deluxetable*}
\setlength{\tabcolsep}{6pt}

\setlength{\tabcolsep}{4pt}
\begin{deluxetable}{lccc}
\tablewidth{\columnwidth}
\tablecaption{Adopted Calibrator Uniform Disk Angular Diameters \label{tab:chara_cal}}
\tablehead{
\colhead{Calibrator}   &
\colhead{$\phi_H$ (mas)} &
\colhead{$\phi_K$ (mas)} &
\colhead{$\sigma_{\phi}$ (mas)} 
}
\startdata
HD 91953    & 0.415   & 0.417  & 0.010 \\
HD 98219    & 0.327   & 0.328  & 0.008 \\
HD 98729    & 0.423   & 0.425  & 0.010 \\
HD 100110   & 0.387   & 0.388  & 0.009 \\
HD 103155   & 0.555   & 0.558  & 0.013 \\
HD 103580   & 0.455   & 0.457  & 0.011 \\  
HD 105395   & 0.373   & 0.374  & 0.009 \\
HD 107662   & 0.463   & 0.465  & 0.012 \\
HD 135969   & 0.215   & 0.216  & 0.006 \\
HD 139487   & 0.310   & 0.311  & 0.007 \\
HD 141162   & 0.476   & 0.478  & 0.011 \\
HD 145375   & 0.492   & 0.494  & 0.012 \\
HD 145965   & 0.212   & 0.213  & 0.005 \\
HD 151259   & 0.425   & 0.427  & 0.011 \\
HD 152557   & 0.382   & 0.384  & 0.009 \\
HD 153213   & 0.260   & 0.262  & 0.006 \\
HD 156365   & 0.408   & 0.410  & 0.010 \\
HD 163843   & 0.375   & 0.377  & 0.009 \\
HD 169580   & 0.352   & 0.354  & 0.008 \\
HD 170347   & 0.390   & 0.392  & 0.009 \\
HD 175120   & 0.318   & 0.319  & 0.008 \\
HD 178834   & 0.421   & 0.423  & 0.010 
\enddata
\end{deluxetable}
\setlength{\tabcolsep}{6pt}

\setlength{\tabcolsep}{6pt}
\begin{deluxetable*}{llcccccl}
\tablewidth{\textwidth}
\tablecaption{New and Archival Interferometric Measurements for \hd\ \label{tab:interf}}
\tablehead{
\colhead{UT Date} &
\colhead{MJD} &
\colhead{$\Delta\alpha$ (mas)} &
\colhead{$\Delta\delta$ (mas)} &
\colhead{Flux ratio} &
\colhead{$f_{\rm incoh}$} &
\colhead{Band} &
\colhead{Instrument}
}
\startdata
\multicolumn{8}{c}{\hda\ (Aa--Ab relative positions)} \\ [0.5ex]
\hline \\ [-1.5ex]
2021 May 11  &  59345.185  &  $-3.03 \pm 0.02$\phs  &  $-11.98 \pm 0.04$\phn\phs  &  $0.147 \pm 0.009$  &  0.47   &  $H$  &  CHARA/MIRC-X \\         
2021 May 12  &  59346.160  &  $-2.65 \pm 0.06$\phs  &  $-12.53 \pm 0.06$\phn\phs  &  $0.151 \pm 0.014$  &  0.60   &  $H$  &  CHARA/MIRC-X \\         
2021 May 14  &  59348.178  &  $-1.95 \pm 0.02$\phs  &  $-13.41 \pm 0.03$\phn\phs  &  $0.119 \pm 0.011$  &  0.51   &  $H$  &  CHARA/MIRC-X \\         
2023 Apr 27  &  60061.235  &  $11.01 \pm 0.03$\phn  &   $12.37 \pm 0.01$\phn      &  $0.113 \pm 0.007$  &  0.52   &  $H$  &  CHARA/MIRC-X \\         
2023 Apr 28  &  60062.260  &  $10.59 \pm 0.03$\phn  &   $12.80 \pm 0.05$\phn      &  $0.124 \pm 0.011$  &  0.57   &  $H$  &  CHARA/MIRC-X \\         
2024 May 28  &  60458.206  &  $13.63 \pm 0.03$\phn  &  $-17.77 \pm 0.04$\phn\phs  &  $0.096 \pm 0.004$  & \nodata &  $H$  &  CHARA/MIRC-X \\         
2024 May 28  &  60458.206  &  $13.61 \pm 0.03$\phn  &  $-17.96 \pm 0.04$\phn\phs  &  $0.146 \pm 0.006$  & \nodata &  $K$  &  CHARA/MYSTIC \\         
2024 Jun 6--7 & 60468.001  &  $15.67 \pm 0.11$\phn  &  $-16.67 \pm 0.08$\phn\phs  &  $0.128 \pm 0.020$  &  0.45   &  $H$  &  VLTI/PIONIER \\         
2024 Jun 20  &  60481.980  &  $18.66 \pm 0.29$\phn  &  $-14.57 \pm 0.23$\phn\phs  &  $0.118 \pm 0.007$  & \nodata &  $H$  &  VLTI/PIONIER \\         
2025 May  3  &  60798.232  &  $20.27 \pm 0.12$\phn  &   $-1.29 \pm 0.06$\phs      &  $0.118 \pm 0.014$  &  0.53   &  $H$  &  CHARA/MIRC-X \\         
2025 May  3  &  60798.232  &  $20.07 \pm 0.07$\phn  &   $-1.64 \pm 0.07$\phs      &  $0.178 \pm 0.012$  &  0.50   &  $K$  &  CHARA/MYSTIC \\ [0.5ex] 
\hline \\ [-1.5ex]
\multicolumn{8}{c}{\hdb\ (Ba--Bb relative positions)} \\ [0.5ex]
\hline \\ [-1.5ex]
2024 May 28  &  60458.206  &  $14.44 \pm 0.03$\phn  &    $3.63 \pm 0.04$          &  $0.548 \pm 0.020$  & \nodata &  $H$  &  CHARA/MIRC-X \\         
2024 May 28  &  60458.206  &  $14.49 \pm 0.03$\phn  &    $3.62 \pm 0.07$          &  $0.474 \pm 0.018$  & \nodata &  $K$  &  CHARA/MYSTIC \\         
2024 Jun 20  &  60481.980  &  $16.15 \pm 0.16$\phn  &    $0.25 \pm 0.18$          &  $0.557 \pm 0.030$  & \nodata &  $H$  &  VLTI/PIONIER \\         
2026 Mar 25  &  61124.353  &  $17.15 \pm 0.02$\phn  &   $-2.11 \pm 0.03$\phs      &  $0.745 \pm 0.019$  & \nodata &  $H$  &  CHARA/MIRC-X \\         
2026 Mar 25  &  61124.353  &  $17.19 \pm 0.01$\phn  &   $-1.97 \pm 0.01$\phs      &  $0.736 \pm 0.010$  & \nodata &  $K$  &  CHARA/MYSTIC \\ [0.5ex] 
\hline \\ [-1.5ex]
\multicolumn{8}{c}{\hd\ AB (Aa--Ba relative positions)} \\ [0.5ex]
\hline \\ [-1.5ex]
2024 May 28  &  60458.206  &  $33.82 \pm 0.01$\phn  &  $195.00 \pm 0.02$          &        0.41         & \nodata &  $H$  &  CHARA/MIRC-X \\         
2024 May 28  &  60458.206  &  $34.03 \pm 0.02$\phn  &  $195.08 \pm 0.07$          &        0.71         & \nodata &  $K$  &  CHARA/MYSTIC \\         
2024 Jun 20  &  60481.980  &  $34.38 \pm 0.13$\phn  &  $195.07 \pm 0.09$          &        0.42         & \nodata &  $H$  &  VLTI/PIONIER \\         
2026 Mar 25  &  61124.353  &  $26.37 \pm 0.01$\phn  &  $120.45 \pm 0.02$          &        0.78         & \nodata &  $H$  &  CHARA/MIRC-X \\         
2026 Mar 25  &  61124.353  &  $26.44 \pm 0.01$\phn  &  $120.62 \pm 0.01$          &        0.86         & \nodata &  $K$  &  CHARA/MYSTIC   
\enddata
\tablecomments{For observations showing only one of the visual components (\hda\ or B),
the model included a contribution from incoherent flux ($f_{\rm incoh}$) due to the other component. The Aa-Ba flux ratios in the bottom section are given without
uncertainties, as they are not particularly meaningful because they can be
significantly biased by bandwidth smearing in cases where the field of
view is limited due to lower spectral resolution. Position uncertainties in the table are statistical only. For example, adjustments made in the course of our orbital analysis below led to more realistic median errors of about 0.09\,mas for binary A and 0.3\,mas for B in each coordinate.}
\end{deluxetable*}

\subsubsection{VLTI}
\label{sec:vlti}

The interferometric observations of \cite{Zuniga:2021} with which they spatially resolved the A component of \hd\ for the first time were obtained with the Very Large Telescope Interferometer (VLTI) and its PIONIER beam combiner, and were published in the form of relative positions at 8 epochs between 2019 and 2021. The observations were made in the $H$ band, using the 1.8\,m Auxiliary Telescopes. We used them here as published. The same authors also reported two observations of the B component collected in a similar manner, and one more was later added by \cite{Zuniga:2026}. We incorporated these three observations of B into our analysis as well.

Two other calibrated interferometric observations of \hd\ from the  VLTI that we found to be useful for our purposes are archived at the Optical interferometry Data Base (OiDB)\footnote{\url{https://oidb.jmmc.fr/index.html}}, maintained
at the Jean-Marie Mariotti Center. A summary of these observations is given in
Table~\ref{tab:archival_vlti}. {They were obtained with the same instrumental setup as the previously published VLTI observations, also using a spectral resolution of $R = 34$, and were calibrated as described by \cite{LeBouquin:2011}.}

\setlength{\tabcolsep}{2pt}
\begin{deluxetable}{lccccc}
\tablewidth{\columnwidth}
\tablecaption{Archival VLTI Observations for \hd\ \label{tab:archival_vlti}}
\tablehead{
\colhead{UT Date} &
\colhead{MJD} &
\colhead{Program ID} &
\colhead{Baselines} &
\colhead{Seeing}
}
\startdata
2024 Jun 6--7  &  60468.001  &  113.26ML.006  &  A0--B5--J2--J6  &  0.57 \\
2024 Jun 20    &  60481.980  &  113.26ML.001  &  A0--B2--D0--C1  &  2.1 
\enddata
\tablecomments{The seeing given in the last column (in arc seconds) is
the average over the observation.}
\end{deluxetable}

\subsubsection{CHARA and VLTI Measurements}
\label{sec:charavlti}

The calibrated squared visibilities (fringe contrast) and closure phases from each of the CHARA and VLTI observations of \hd\ were analyzed simultaneously to infer the relative positions for each night, adopting a model of either two or four stars, depending on the epoch and instrument configuration (spectral resolution, wavelength, baselines). Our earlier observations do not allow us to measure all four stars,
while some of the later ones do.
This is a function of both the angular separation between A and B, which changed between 0\farcs34 and 0\farcs12 from 2021 to 2026, and the effective field of view (FoV). The interferometric FoV is set approximately by the product of the spectral resolution ($R = \lambda/\Delta\lambda$) and the fringe angular scale ($\lambda/B$), where $B$ is the baseline. For the CHARA observations of \hd, the baselines ranged from about 10\,m to 270\,m, while for the VLTI observations they were $\sim$50--200\,m (UT 2024\,Jun\,6--7) and $\sim$11--34\,m (UT 2024\,Jun\,20).

Binary A was clearly resolved in our early data, and aside from the relative position ($\Delta\alpha$, $\Delta\delta$) and flux ratio, our modeling included an additional parameter to account for the contribution from the incoherent flux ($f_{\rm incoh}$) of binary B. On UT 2024\,May\,28 (CHARA) and UT 2024\,Jun\,20 (VLTI), both binaries were measurable, and we analyzed the observation with a quadruple star model, solving for the positions of stars Ab, Ba, and Bb relative to Aa. We also determined the flux ratios for the inner binaries, and for the outer pair. However, the latter is affected by bandwidth smearing, so we do not consider it an accurate measure of the flux ratio B/A. There may also be some effect on the ratios for the inner binaries.

By the time of our CHARA observation made on UT 2026\,Mar\,25, \hda\ had already moved behind the circumbinary disk surrounding B, and was being subjected to considerable extinction (see Section~\ref{sec:eclipseA}), such that star Ab was not detected. In this case, we fitted a model of only three stars.

The apparent sizes of the four components are too small to be resolved by the CHARA and VLTI observations. Their predicted diameters, based on the mass estimates resulting from our analysis together with stellar evolution models described later, are 0.23, 0.16, 0.21, and 0.21\,mas for components Aa, Ab, Ba, and Bb, respectively. Our modeling above included the diameters as adjustable variables, merely as a device for taking into account their formal uncertainties ($\sim$0.01\,mas). This was done by applying Gaussian priors to those four quantities in our Markov chain Monte Carlo procedures.
Our modeling also accounted for bandwidth smearing following \cite{Gardner:2021}. For the CHARA observations, we applied a recommended correction to the wavelengths, reducing them by factors of $1.0054 \pm 0.0006$ for MIRC-X and $1.0067 \pm 0.0007$ for MYSTIC, following \cite{Gardner:2022} and Gardner (2022, private communication). For the MIRC-X observations collected in 2025, the correction factor  was slightly different ($0.999 \pm 0.001$)\footnote{Data Reduction Manual for MIRC-X and MYSTIC, version 0.9.6, \url{https://chara.gsu.edu/public/instrumentation/mirc}}, leading to an upward adjustment to the wavelengths.
The results from the measurement of these interferometric observations are collected in Table~\ref{tab:interf}, where the uncertainties are statistical only, and do not account for possible systematics. More realistic uncertainties are derived in Section~\ref{sec:orbit}.

\subsection{Astrometric Observations From the Literature}
\label{sec:WDS}

Since its discovery as a visual binary by \cite{Innes:1909}, \hd\ has
been measured nearly two dozen times by the visual observers, and more
recently also with the speckle interferometry technique. These
observations are essential for constraining the outer orbit of the
system, which has a $\sim$240~yr period. A listing of 48
measurements (1909--2023)\footnote{A few of the measurements in the original listing were excluded for the following reasons. One by \cite{Torres:2006} (no relation to the present author), from 2002, is not an independent measurement. Another by \cite{Tokovinin:2020}, from 2019, was duplicated. Another by \cite{Zuniga:2021}, from 2021, was superseded in a later paper \citep{Tokovinin:2022}. Finally, a 2017 speckle observation of the visual pair by \cite{Horch:2019}, made at two different wavelengths, was the first to also resolve the B binary with this technique. However, the Ba--Bb separation was actually below the diffraction limit of the telescope, and the measurements for both Aa--Ba and Ba--Bb were found to deviate significantly from
the expected positions. The reduction of those difficult observations had originally been carried out with an experimental procedure that has since been abandoned. While a re-reduction with standard methods yielded some improvement, we have nevertheless considered it safer to omit these measurements from the present orbital analysis.}, as contained in the Washington Double Star Catalog \citep[WDS;][]{Worley:1997, Mason:2001}, was supplemented with two additional adaptive optics measurements by \cite{Koerner:2000} and \cite{Prato:2001}, and additional speckle measurements obtained more recently by \cite{Tokovinin:2026}. {For the benefit of the reader, we collect all of these measurements in Table~\ref{tab:wds}.} Further speckle observations were gathered on 2026 March 1st with the Differential Speckle Survey Instrument (DSSI) on the Astrophysical Research Consortium 3.5\,m telescope at Apache Point Observatory (APO) \cite[see][]{Davidson:2024}. We list the new DSSI measurements in Table~\ref{tab:speckle}.
Except for the speckle observations, the WDS
measurements prior to 2009 have usually been published without corresponding
uncertainties. To assign them errors, we have generally followed the
prescription of \cite{Douglass:1992}, which relies on an analysis of the residuals of a large number of historical observations from well-studied astrometric orbits {(see Table~\ref{tab:wds}).}

\setlength{\tabcolsep}{3pt}
\begin{deluxetable}{lccccc}
\tablewidth{\columnwidth}
\tablecaption{Visual and Speckle Observations of \hd\ \label{tab:wds}}
\tablehead{
\colhead{Julian Year} &
\colhead{$\theta$} &
\colhead{$\sigma_{\theta}$} &
\colhead{$\rho$} &
\colhead{$\sigma_{\rho}$} &
\colhead{$\lambda$} 
\\
\colhead{} &
\colhead{(degree)} &
\colhead{(degree)} &
\colhead{(\arcsec)} &
\colhead{(\arcsec)} &
\colhead{(nm)}
}
\startdata
1909.5   &  190\phm{.0}  &  2.9  &  1.0\phn  &  0.09  &  550 \\
1910.3   &  180\phm{.0}  &  3.6  &  0.8\phn  &  0.09  &  550 \\
1912.66  &  187.0        &  4.4  &  0.65     &  0.09  &  550 \\
1926.25  &  192.5        &  5.6  &  0.41     &  0.07  &  550 \\
1930.39  &  194.0        &  6.2  &  0.37     &  0.07  &  550 
\enddata

\tablecomments{The source for the majority of these relative positions
  is the WDS, where the original bibliographic references can be
  found.  Wavelengths $\lambda$ prior to 1990, not specified in the
  WDS, were assumed to be 550\,nm. The 1998.197 observation by
  \cite{Koerner:2000} consists only of an angular separation (no
  position angle was reported). The 1998.356 measurement by
  \cite{Prato:2001} corresponds strictly to a north-south
  separation. (This table is available in its entirety in
  machine-readable form.)}

\end{deluxetable}

\setlength{\tabcolsep}{2pt}
\begin{deluxetable}{cccccc}
\tablewidth{\columnwidth}
\tablecaption{New DSSI Speckle Observations of \hd\ \label{tab:speckle}}
\tablehead{
\colhead{Julian Year} &
\colhead{$\theta$} &
\colhead{$\rho$} &
\colhead{$\Delta m$} &
\colhead{$\lambda$} &
\colhead{Seeing}
\\
\colhead{} &
\colhead{(degree)} &
\colhead{(\arcsec)} &
\colhead{(mag)} &
\colhead{(nm)} &
\colhead{(\arcsec)}
}
\startdata
2026.1624  &  $15.6 \pm 0.8$\phn & $0.1277 \pm 0.0035$ & 0.37 & 692 & 0.61 \\
2026.1624  &  $15.9 \pm 0.8$\phn & $0.1258 \pm 0.0035$ & $-0.01$\phs & 880 & 0.60 
\enddata
\end{deluxetable}

In their analysis of the B binary, \cite{Boden:2005} reported measurements from the Hubble Space Telescope (HST) Fine Guidance Sensors (FGS) that resolved the components, although the separations were near the resolution limit of the FGS. Unfortunately, these measurements are of much lower precision compared to the KI, CHARA, and VLTI measurements, and we have chosen not to consider them in our analysis, as they do not contribute significantly.


\section{Orbital Solution}
\label{sec:orbit}

Our approach in this work has been to combine all observations of different
types into a global solution, to determine the elements of the inner
and outer orbits simultaneously, along with other derived properties.
This has the advantage of self-consistency, and maximizes the use of
information by permitting the observations to constrain more
parameters at the same time. It also leads to more accurate
uncertainties that properly account for correlations among the
elements. Furthermore, it becomes possible to exploit redundancies
among the elements that can reduce the number of free parameters, as
we show below.  This differs from the piecemeal approaches adopted in
earlier studies,
in which
results from separate fits to the inner orbits were subsequently used
to constrain the outer orbit, sometimes aided by external information
such as a trigonometric parallax.

Traditionally, the astrometric elements of the outer orbit needed to
model the observations are $P_{\rm AB}$, $a^{\prime\prime}_{\rm AB}$,
$e_{\rm AB}$, $i_{\rm AB}$, $\omega_{\rm A}$, $\Omega_{\rm AB}$, and
$T_{\rm peri,AB}$, in which the symbols have their usual meaning, and
where $\omega_{\rm A}$ corresponds to the argument of periastron for
the primary (\hda). For numerical efficiency, the latter angle and the
eccentricity have been recast here as $\sqrt{e_{\rm AB}}\cos\omega_{\rm A}$ and
$\sqrt{e_{\rm AB}}\sin\omega_{\rm A}$, and the inclination angle as
$\cos i_{\rm AB}$. The three additional spectroscopic
elements needed for modeling the RVs are the
velocity semiamplitudes of the {centers of mass of the} A and B binaries in the outer orbit,
$K_{\rm A}$ and $K_{\rm B}$,
and finally $\gamma$, the center-of-mass velocity of the quadruple
system. The inner orbits of the A and B binaries can be represented by
a similar set of astrometric and spectroscopic elements, except that
the semiamplitude $K_{\rm Ab}$ is not required because binary A is
single-lined.

\begin{figure}
\includegraphics[width=\columnwidth]{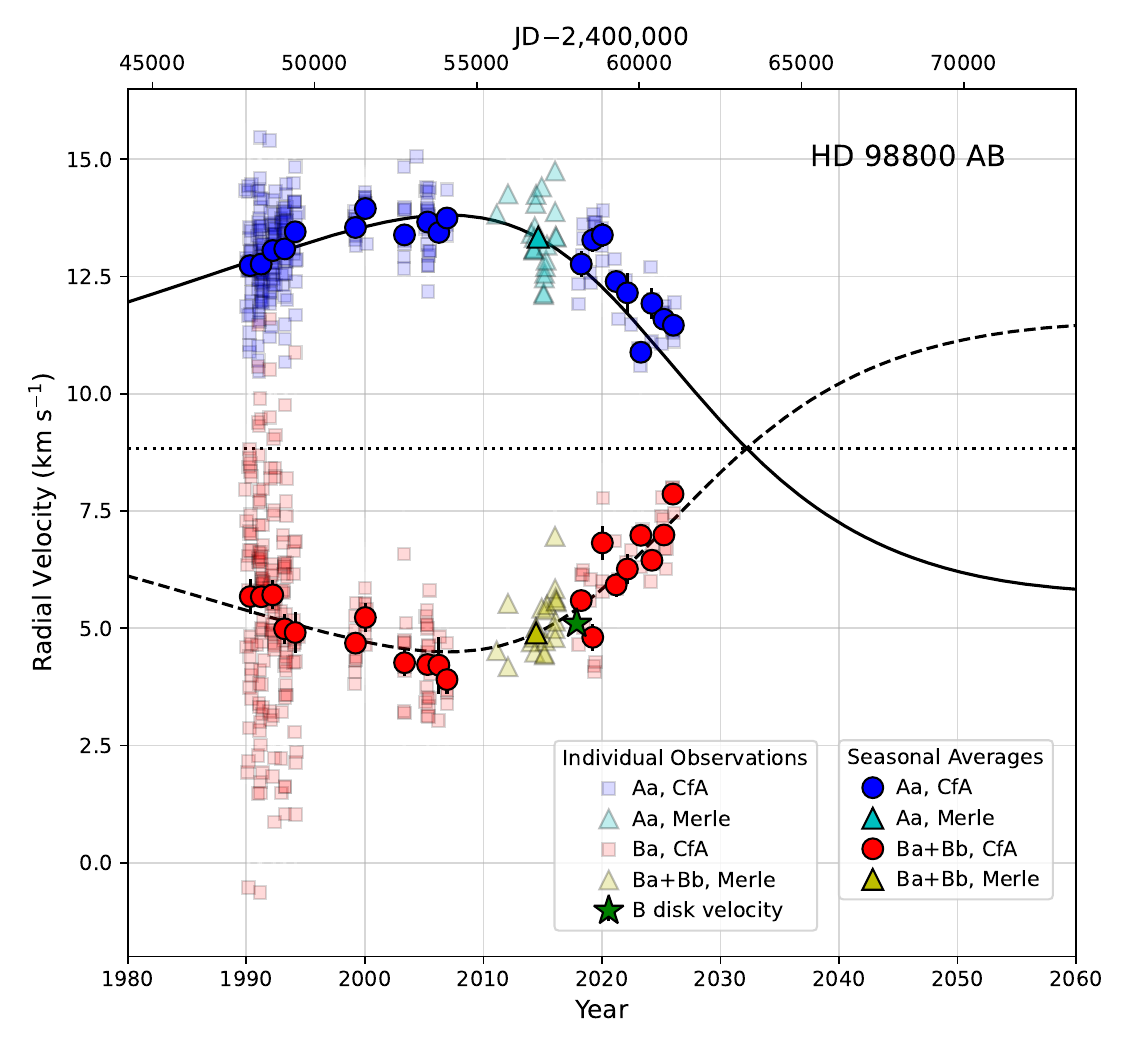}
\figcaption{RVs of \hd\ in the outer orbit. Light-colored symbols
represent individual observations, after removing the short-period
motions in the inner orbits. Darker symbols correspond to seasonal
averages, shown here only to guide the eye. They were not used in
our solution. The curves represent our model for the outer orbit,
and the dotted line marks the center-of-mass velocity of the
quadruple system.\label{fig:RVout}}
\end{figure}

The CfA RVs of \hd\ span more than three decades, and by now the velocities of
A and B show obvious drifts in opposite directions caused by motion in
the outer orbit, superimposed on the shorter-term variations within
each binary.  The AB pair is then effectively a double-lined
spectroscopic binary, and both $K_{\rm A}$ and $K_{\rm B}$ can be
constrained to some extent by the observations.  We are fortunate that
the RVs happen to sample the nodal passage in the outer orbit, as
illustrated in Figure~\ref{fig:RVout}, providing maximal leverage for
determining the velocity semiamplitudes. 

In principle, knowledge of the astrometric and spectroscopic elements
of AB allows us to directly infer the total masses of the A and B
binaries, along with the orbital parallax for the system.  In practice, however, the
phase coverage in the outer orbit is far from complete, and it becomes
necessary to supplement the RVs with the AB astrometric observations
in order to constrain the velocity semiamplitudes sufficiently well.
The B binary, which is itself also double-lined, offers the
possibility of deriving another value for the orbital parallax, in
addition to the masses of the individual components. These two values
of the parallax must be equal. Here we used this redundancy to our
advantage, in order to eliminate from the analysis one of the more
poorly constrained elements of the outer orbit, $a^{\prime\prime}_{\rm
  AB}$. This was done as follows. The orbital parallaxes of B and AB are expressed as
\begin{align}
\pi_{\rm B}  &= {\rm const} \times \frac{a^{\prime\prime}_{\rm B} \sin i_{\rm B} } {P_{\rm B} \sqrt{1-e_{\rm B}^2} (K_{\rm Ba} + K_{\rm Bb}) } \label{eq:piB} \\
\pi_{\rm AB} &= {\rm const} \times \frac{a^{\prime\prime}_{\rm AB} \sin i_{\rm AB} } {P_{\rm AB} \sqrt{1-e_{\rm AB}^2} (K_{\rm A} + K_{\rm B}) },
\end{align}
with the numerical constant being equal to 10879.1 when the periods
are in units of days, the angular semimajor axes in mas, and the
velocity semiamplitudes in \kms. Equating these two expressions leads
to a formula for the semimajor axis of AB,
\begin{equation}
a^{\prime\prime}_{\rm AB} = a^{\prime\prime}_{\rm B} \frac{\sin i_{\rm B}}{\sin i_{\rm AB}} \frac{P_{\rm AB}}{P_{\rm B}} \sqrt{\frac{1-e_{\rm AB}^2}{1-e_{\rm B}^2}} \frac{K_{\rm A} + K_{\rm B}}{K_{\rm Ba} + K_{\rm Bb}},\label{eq:aAB}
\end{equation}
which can then be evaluated using other known elements.

It is also possible to derive expressions for the more poorly constrained
elements $K_{\rm A}$ and $K_{\rm B}$ in terms of other known
quantities, as we now explain.
We begin by noting that the total mass of binary A can be calculated
in two ways, using the elements of A itself, or those of the outer
orbit:
\begin{align}
M_{\rm A}^{\rm in}  &= {\rm const} \times \frac{P_{\rm A} (1-e_{\rm A}^2)^3 (K_{\rm Aa}+K_{\rm Ab})^3}{\sin i_{\rm A}^3} \label{eq:MAin}\\
M_{\rm A}^{\rm out} &= {\rm const} \times \frac{P_{\rm AB} (1-e_{\rm AB}^2)^3 (K_{\rm A}+K_{\rm B})^2 K_{\rm B}}{\sin i_{\rm AB}^3},
\end{align}
where the superscripts indicate whether the inner or outer elements
are involved. In eq.[\ref{eq:MAin}], the parameter $K_{\rm Ab}$ cannot be measured
directly because binary A is single-lined. However it is possible to
derive an expression for it (or more conveniently, for the $K_{\rm
Aa}+K_{\rm Ab}$ sum) by enforcing $\pi_{\rm A} = \pi_{\rm B}$, with the
expression for $\pi_{\rm A}$ being analogous to eq.[\ref{eq:piB}]. This yields
\begin{equation}
K_{\rm Aa} + K_{\rm Ab} = \frac{a^{\prime\prime}_{\rm A}}{a^{\prime\prime}_{\rm B}} \frac{\sin i_{\rm A}}{\sin i_{\rm B}} \frac{P_{\rm B}}{P_{\rm A}} \sqrt{\frac{1-e_{\rm B}^2}{1-e_{\rm A}^2}} (K_{\rm Ba}+K_{\rm Bb}).
\end{equation}
On the other hand, the total mass of B can also be calculated in two
ways, using the elements of B or those of the outer orbit:
\begin{align}
M_{\rm B}^{\rm in}  &= {\rm const} \times \frac{P_{\rm B} (1-e_{\rm B}^2)^3 (K_{\rm Ba}+K_{\rm Bb})^3}{\sin i_{\rm B}^3} \\
M_{\rm B}^{\rm out} &= {\rm const} \times \frac{P_{\rm AB} (1-e_{\rm AB}^2)^3 (K_{\rm A}+K_{\rm B})^2 K_{\rm A}}{\sin i_{\rm AB}^3}.
\end{align}

If we now set $M_{\rm A}^{\rm in} = M_{\rm A}^{\rm out}$ and $M_{\rm
  B}^{\rm in} = M_{\rm B}^{\rm out}$, these two equations allow us to
solve for $K_{\rm A}$ and $K_{\rm B}$. After some algebra, they can be
expressed as $K_{\rm A} = y/(x+y)^{2/3}$ and $K_{\rm B} =
x/(x+y)^{2/3}$, where
\begin{align}
x &= \alpha \left(\frac{P_{\rm B}}{P_{\rm A}}\right)^2 \! \left(\frac{a^{\prime\prime}_{\rm A}}{a^{\prime\prime}_{\rm B}}\right)^3 \\
y &= \alpha
\end{align}
and
\begin{equation}
\alpha = \frac{P_{\rm B}}{P_{\rm AB}} \left(\frac{1-e_{\rm B}^2}{1-e_{\rm AB}}\right)^{3/2} \! \left(\frac{\sin i_{\rm AB}}{\sin i_{\rm B}}\right)^3 (K_{\rm Ba}+K_{\rm Bb})^3.
\end{equation}

The expressions for $K_{\rm A}$ and $K_{\rm B}$ can then be inserted
back into eq.[\ref{eq:aAB}], and in this way the outer elements
$a^{\prime\prime}_{\rm AB}$, $K_{\rm A}$, and $K_{\rm B}$ can all be
put in terms of known elements of the inner orbits, together with the period,
eccentricity, and inclination angle of AB, and they can be removed as free
parameters in the analysis.

The solution of the remaining set of 24 orbital elements, along with others added below, was performed
in a Markov chain Monte Carlo (MCMC) framework using the {\sc emcee}
package\footnote{\url{https://emcee.readthedocs.io/en/stable/index.html}}
of \cite{Foreman-Mackey:2013}. Position angles from the WDS were
precessed to J2000, and although small, light travel time corrections
were nevertheless applied to the timestamps of the observations of \hda\
and B for completeness, and iterated during the analysis.  Strictly,
then, the times of periastron passage $T_{\rm peri,A}$ and $T_{\rm
peri,B}$ are referred to the center of mass of the quadruple system.
For a properly balanced solution, the weighting of each kind of
observation was adjusted iteratively in order to achieve separate reduced $\chi^2$
values near unity. This was done by applying appropriate scaling
factors to the internal errors reported earlier. We found, for example, that the formal uncertainties for the CHARA and VLTI measurements of the inner binaries reported in Table~\ref{tab:interf} are underestimated by factors of roughly 2.4 for binary A and 15 for B. The RV errors for the CfA measurements were scaled by factors ranging from 0.4 to 1.0, depending on the component and instrument (DS or TRES), and the WDS measurements errors in position angle and separation were both increased by a factor of 1.7.

In our procedures for deriving the CfA RVs, star Bb is cool enough
that the synthetic template we used for that component likely does not
match the spectrum perfectly, as model atmospheres for such cool stars still struggle
to reproduce their line profiles at high resolution. This can introduce a systematic shift in
the velocities of Bb relative to those of the other two stars. To account
for this possibility, we introduced two additional free parameters to
represent those offsets, one for the DS measurements ($\Delta$RV$_{\rm
  DS,Bb}$) and another for TRES ($\Delta$RV$_{\rm TRES,Bb}$). The
results reported below confirm our suspicions, as both offsets turn
out to be statistically significant. For the \cite{Merle:2024} RVs, we
solved for a global offset ($\Delta$RV$_{\rm Merle}$) to correct for a
possible velocity zero-point difference compared to the CfA RVs, as
well as a possible shift between stars Ba and Bb
($\Delta$RV$_{\rm Merle,Ba}$), which those authors claim to detect in
their own analysis.

The use of the KI visibilities for A and B requires solving for {an additional} two parameters, representing the $K$-band flux ratios between the primary and secondary in each binary, $f_{\rm A}$ and $f_{\rm B}$. The total number of free parameters is then 30. The angular diameters adopted for the analysis of the KI data are those given in Section~\ref{sec:charavlti}.

As mentioned earlier, initial solutions revealed significant tension between the full set of RVs and the recalibrated KI measurements for \hdb, such that the residuals of the visibilities from the global solution were larger than we expected, and were systematic on several nights. Hints of this were already apparent in remarks made by \cite{Zuniga:2021}. Those authors combined their VLTI measurements for B with the original KI data from \cite{Boden:2005} and available RVs, and noted that their results for some of the parameters of the B orbit seemed inconsistent with those of \cite{Boden:2005}. Details of any relative weighting of the observations that may have been applied were not provided, but the authors did suggest that some of the uncertainties may have been underestimated. More recently, \cite{Zuniga:2026} again incorporated those original visibility measurements into their updated solution for \hdb, and chose to reweight the KI data by adding a jitter term in quadrature to the internal $V^2$ errors, separately for each night.

Given that it is clear the KI measurements do contain useful information on the orbital motion worth preserving, here we have chosen to follow the same prescription, adding a jitter term quadratically to the uncertainties generated by the calibration pipeline, in order to reach reduced $\chi^2$ values near unity on each night. This was done simultaneously with the adjustments applied to the errors of the rest of the observations (see above). While the source of this extra noise for KI has not been identified, it is likely due at least in part to instrumental systematics or to atmospheric effects, both of which can change from night to night. An additional possibility is that the shifts, such as seen particularly on MJD 53101, are of astrophysical origin, i.e., caused by real changes in the orbits. We explore this in more detail in Section~\ref{sec:dynamics}, although our conclusion there will be that there is no compelling evidence of such changes at a level that would make a difference.

\begin{figure}
\includegraphics[width=0.47\textwidth]{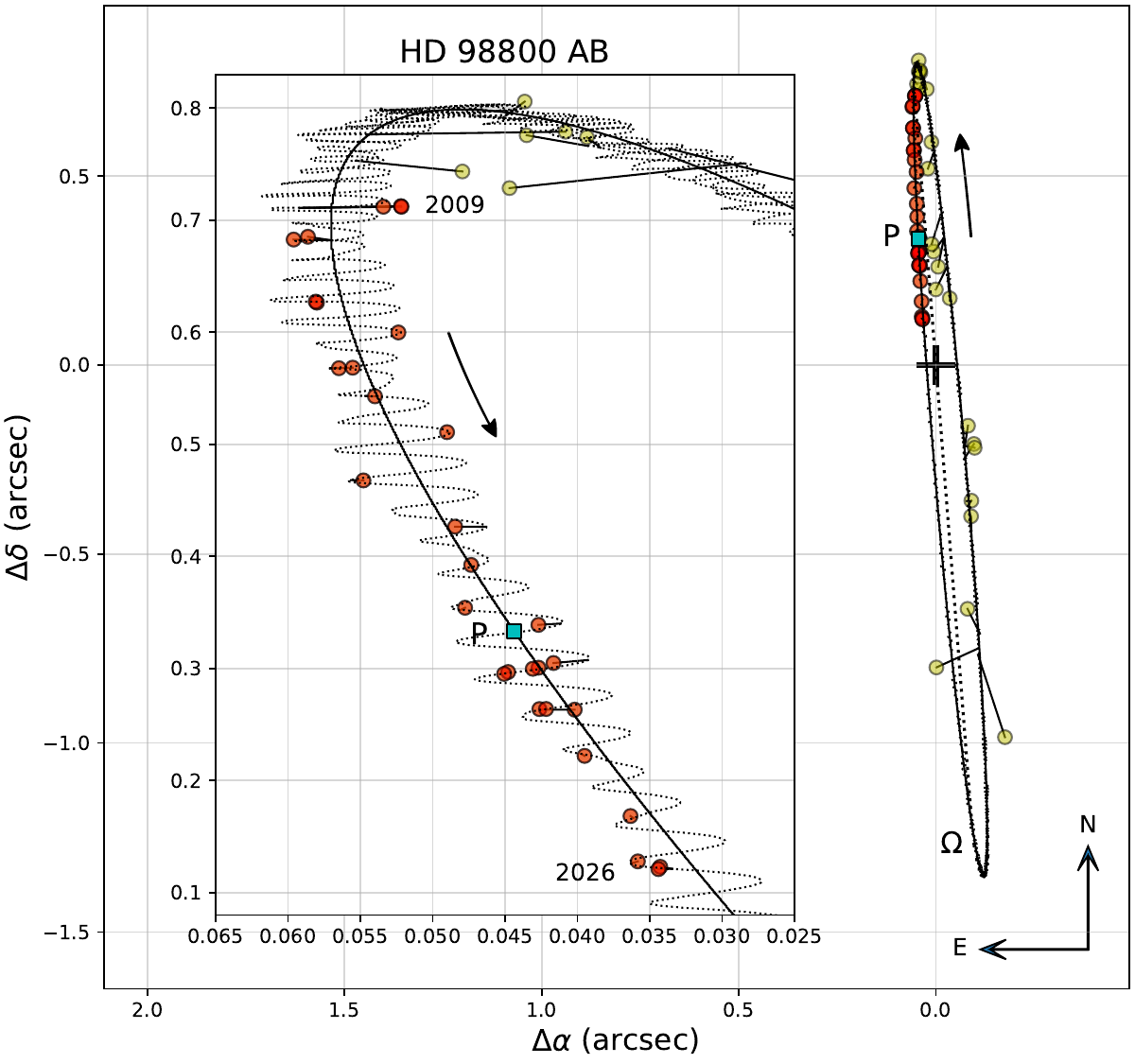}
\figcaption{Observations of the nearly edge-on outer orbit of \hd\ on the plane of the sky, together with our model.
The full orbit is shown on the right, with component A at the origin.
The cyan square symbol marked ``P'' corresponds to periastron,
and the straight dotted line is the line of nodes, the southern end
of which corresponds to the
ascending node, labeled ``$\Omega$''.
Red symbols represent the speckle observations since 2009, made
mostly by A.\ Tokovinin and collaborators. Yellow symbols represent
older measurements from the WDS.
The inset on the left displays an expanded view of the most recent
observations, with the horizontal axis stretched to show the fit
to the observations in greater detail. The solid curve corresponds
to our model for the motion of the center of mass of \hdb\ relative to that of A.
The dotted line represents our full model that accounts for the motion of the center of
light of B relative to the center of light of A, which is effectively what the recent speckle observations actually measure. This model is seen to fit the
measurements well.\label{fig:ABsky}}
\end{figure}

As pointed out by \cite{Zuniga:2021}, the speckle observations of the
AB pair since 2009 are precise enough to reveal the astrometric
wobble caused by the motion in the inner orbits of A and B. Almost all of those measurements were made by A.\ Tokovinin and collaborators, with the
Southern Astrophysical Research (SOAR) 4.1\,m telescope, in Chile.
These observations generally do not resolve either of the inner binaries.
Therefore, rather than measuring the relative position of the centers of mass,
they correspond strictly to the position of the center of
light of \hdb\ relative to the center of light of \hda. 
In our analysis we have accounted for these extra
motions at each epoch of observation by calculating the offsets of the photocenters of A and B relative to their respective centers of mass, based on estimates of the magnitude difference $\Delta m$ between the primary and secondary in each system.
The semimajor axis of the photocenter of A follows from the expression
\begin{equation}
  \label{eq:photocenter}
a^{\prime\prime}_{\rm phot,A} = a^{\prime\prime}_{\rm A} \left(\frac{K_{\rm Aa}}{K_{\rm Aa}+K_{\rm Ab}} - \frac{1}{1+10^{0.4 \Delta m}}\right),
\end{equation}
and an analogous formula holds for B. The resulting wobble effect is significant
(several mas) and complex, and is illustrated in Figure~\ref{fig:ABsky}, where it is seen that the observations do indeed resolve it accurately. 
Because the stars in each inner binary are different,
strictly speaking $\Delta m$ depends on wavelength. We have
approximated that dependence by means of smooth functions drawn
through {flux ratios calculated over the optical and near-infrared range} using appropriate synthetic
spectra for the four components {(see Appendix~A)}. We then used the recorded wavelength of each
observation {(Table~\ref{tab:wds})} to predict $\Delta m$, and to calculate the location of
the centers of light of A and B. 
For example, at a wavelength of $\sim$820\,nm corresponding to the $I$-band filter used in many of the speckle observations, $a^{\prime\prime}_{\rm phot,B}$ is about 2.5\,mas, which is already larger than the formal uncertainties of the more recent measurements. At $\sim$550\,nm ($y$-band filter), $a^{\prime\prime}_{\rm phot,B}$ increases to 4.9\,mas.
For \hda, the center of light is usually close to star Aa because star Ab is much fainter. Star Aa itself moves around the center of mass with a semimajor axis of about 4.4\,mas. At $\sim$550\,nm, the flux from Ab reduces this by only $\sim$0.2\,mas, but at $\sim$820\,nm where the companion is brighter, the change is about 1.2\,mas, which again is not negligible compared to the typical uncertainties.
Although the combined effect of the inner motions of the two binaries is only noticeable in
the speckle measurements, for simplicity we accounted for it in all AB
observations from the WDS, as part of our analysis.

\subsection{Results}
\label{sec:results}

Our global solution for \hd\ gives the orbital elements listed in
Table~\ref{tab:mcmc}.\footnote{
There has been some confusion in the past about the proper quadrant
for the ascending node of the outer orbit, $\Omega_{\rm AB}$. Our
determination is consistent with the conventional definition in which
it is the node at which the secondary (the fainter star in the optical,
i.e., \hdb) is receding from the observer, relative to the center of mass.
The value listed in the table
agrees with those of \cite{Torres:1995}, \cite{Tokovinin:1999},
\cite{Tokovinin:2014}, and \cite{Zuniga:2021, Zuniga:2026}, whereas
\cite{Kennedy:2019} used a different convention, and placed $\Omega_{\rm AB}$ in the opposite quadrant.} Other properties derived from the elements are listed there as well,
including
$a^{\prime\prime}_{\rm AB}$, $K_{\rm A}$, and $K_{\rm B}$, which can be
expressed in terms of other fitted quantities, as described earlier.

\setlength{\tabcolsep}{3pt}
\begin{deluxetable}{lcc}
\tablewidth{\columnwidth}
  \tablecaption{Global Orbital Solution for \hd\ \label{tab:mcmc}}
  \tablehead{
    \colhead{Parameter} &
    \colhead{Value} &
    \colhead{Prior}
  }
  \startdata
  \multicolumn{3}{c}{\hda} \\ [0.5ex]
  \hline \\ [-1.5ex]
  $P_{\rm A}$ (day)                         &  $264.5653 \pm 0.0081$\phn\phn    &  [200, 500] \\
  $T_{\rm peri,A}$ (day)                    &  $49798.88 \pm 0.27$\phm{2222}  &  [49700, 49900] \\
  $K_{\rm Aa}$ (\kms)                       &  $6.628 \pm 0.054$              &  [1, 40] \\
  $a^{\prime\prime}_{\rm A}$ (mas)          &  $19.028 \pm 0.023$\phn         &  [5, 40] \\
  $\sqrt{e_{\rm A}}\cos\omega_{\rm Aa}$     &  $+0.2565 \pm 0.0015$\phs       &  [$-1$, 1] \\
  $\sqrt{e_{\rm A}}\sin\omega_{\rm Aa}$     &  $+0.64527 \pm 0.00092$\phs       &  [$-1$, 1] \\
  $\cos i_{\rm A}$                          &  $-0.7125 \pm 0.0020$\phs       &  [$-1$, 1] \\
  $\Omega_{\rm A}$ (degree)                 &  $169.70 \pm 0.18$\phn\phn      &  [0, 360] \\
  $f_{\rm A}$                               &  $0.1274 \pm 0.0039$            &  [0.01, 1] \\ [+0.5ex]
  \hline \\ [-1.5ex]
  \multicolumn{3}{c}{Derived Properties for \hda} \\ [0.5ex]
  \hline \\ [-1.5ex]
  $e_{\rm A}$                               &  $0.48217 \pm 0.00069$            &  \nodata \\
  $\omega_{\rm Aa}$ (degree)                &  $68.32 \pm 0.14$\phn           &  \nodata \\
  $i_{\rm A}$ (degree)                      &  $135.44 \pm 0.16$\phn\phn      &  \nodata \\
  $K_{\rm Ab}$ (\kms)                       &  $22.11 \pm 0.26$\phn           &  \nodata \\ [+0.5ex]
  \hline \\ [-1.5ex]
  \multicolumn{3}{c}{\hdb} \\ [0.5ex]
  \hline \\ [-1.5ex]
  $P_{\rm B}$ (day)                         &  $314.8125 \pm 0.0062$\phn\phn  &  [200, 500] \\
  $T_{\rm peri,B}$ (day)                    &  $52487.637 \pm 0.061$\phm{2222}  &  [52400, 52600] \\
  $K_{\rm Ba}$ (\kms)                       &  $22.81 \pm 0.13$\phn           &  [1, 40] \\
  $K_{\rm Bb}$ (\kms)                       &  $29.78 \pm 0.21$\phn           &  [1, 40] \\
  $a^{\prime\prime}_{\rm B}$ (mas)          &  $22.366 \pm 0.097$\phn         &  [5, 40] \\
  $\sqrt{e_{\rm B}}\cos\omega_{\rm Ba}$     &  $-0.2336 \pm 0.0026$\phs       &  [$-1$, 1] \\
  $\sqrt{e_{\rm B}}\sin\omega_{\rm Ba}$     &  $+0.8551 \pm 0.0018$\phs       &  [$-1$, 1] \\
  $\cos i_{\rm B}$                          &  $0.3977 \pm 0.0026$            &  [$-1$, 1] \\
  $\Omega_{\rm B}$ (degree)                 &  $342.16 \pm 0.27$\phn\phn      &  [0, 360] \\
  $f_{\rm B}$                               &  $0.6377 \pm 0.0050$            &  [0.01, 1] \\
  $\Delta$RV$_{\rm DS,Bb}$ (\kms)           &  $-1.02 \pm 0.23$\phs           &  [$-$5, 5] \\
  $\Delta$RV$_{\rm TRES,Bb}$ (\kms)         &  $-0.58 \pm 0.24$\phs           &  [$-$5, 5] \\
  $\Delta$RV$_{\rm Merle}$ (\kms)           &  $+0.522 \pm 0.092$\phs         &  [$-$5, 5] \\
  $\Delta$RV$_{\rm Merle,Ba}$ (\kms)        &  $-0.39 \pm 0.18$\phs           &  [$-$5, 5] \\ [+0.5ex]
  \hline \\ [-1.5ex]
  \multicolumn{3}{c}{Derived Properties for \hdb} \\ [0.5ex]
  \hline \\ [-1.5ex]
  $e_{\rm B}$                               &  $0.7857 \pm 0.0021$            &  \nodata \\
  $\omega_{\rm Ba}$ (degree)                &  $105.28 \pm 0.19$\phn\phn      &  \nodata \\
  $i_{\rm B}$ (degree)                      &  $66.56 \pm 0.16$\phn           &  \nodata \\ [+0.5ex]
  \hline \\ [-1.5ex]
  \multicolumn{3}{c}{Outer Orbit (HD 98800 AB)} \\ [0.5ex]
  \hline \\ [-1.5ex]
  $P_{\rm AB}$ (yr)                         &  $238.0 \pm 8.0$\phn\phn        &  [100, 500] \\
  $T_{\rm peri,AB}$ (yr)                    &  $2021.45 \pm 0.40$\phm{222}    &  [1980, 2050] \\
  $\sqrt{e_{\rm AB}}\cos\omega_{\rm A}$     &  $+0.354 \pm 0.020$\phs         &  [$-1$, 1] \\
  $\sqrt{e_{\rm AB}}\sin\omega_{\rm A}$     &  $+0.5730 \pm 0.0041$\phs       &  [$-1$, 1] \\
  $\cos i_{\rm AB}$                         &  $0.03545 \pm 0.00041$          &  [$-1$, 1] \\
  $\Omega_{\rm AB}$ (degree)                &  $184.297 \pm 0.032$\phn\phn    &  [0, 360] \\
  $\gamma$ (\kms)                           &  $8.830 \pm 0.050$              &  [0, 20] \\ [+0.5ex]
  \hline \\ [-1.5ex]
  \multicolumn{3}{c}{Derived Properties for Outer Orbit (HD 98800 AB)} \\ [0.5ex]
  \hline \\ [-1.5ex]
  $e_{\rm AB}$                              &  $0.453 \pm 0.011$              &  \nodata \\
  $\omega_{\rm A}$ (degree)                 &  $58.3 \pm 1.6$\phn             &  \nodata \\
  $i_{\rm AB}$ (degree)                     &  $87.968 \pm 0.023$\phn         &  \nodata \\
  $a^{\prime\prime}_{\rm AB}$ (\arcsec)     &  $1.169 \pm 0.028$              &  \nodata \\
  $K_{\rm A}$ (\kms)                        &  $4.018 \pm 0.025$              &  \nodata \\
  $K_{\rm B}$ (\kms)                        &  $3.503 \pm 0.058$              &  \nodata 
  \enddata
  \tablecomments{The values listed correspond to the median of the posterior distributions,
    with the uncertainties representing the 68.3\% credible intervals. Priors
    were uniform in all cases. The times of
    periastron passage $T_{\rm peri,A}$ and $T_{\rm peri,B}$ are referred to BJD 2,400,000.}

\end{deluxetable}

For the outer orbit, our astrometric model together with the observations was
shown previously in Figure~\ref{fig:ABsky}, and the spectroscopic model with
the data was presented in Figure~\ref{fig:RVout}. The green star-shaped symbol we have added in the latter figure
corresponds to the mean RV of the circumbinary disk around \hdb\ ($5.1 \pm 0.1~\kms$),
as measured by \cite{Kennedy:2019}, which was not used in our solution.
On the assumption that the center of the disk is at the center of mass of
the binary, the good agreement with our model for \hd\ is
reassuring.

\begin{figure}
\includegraphics[width=\columnwidth]{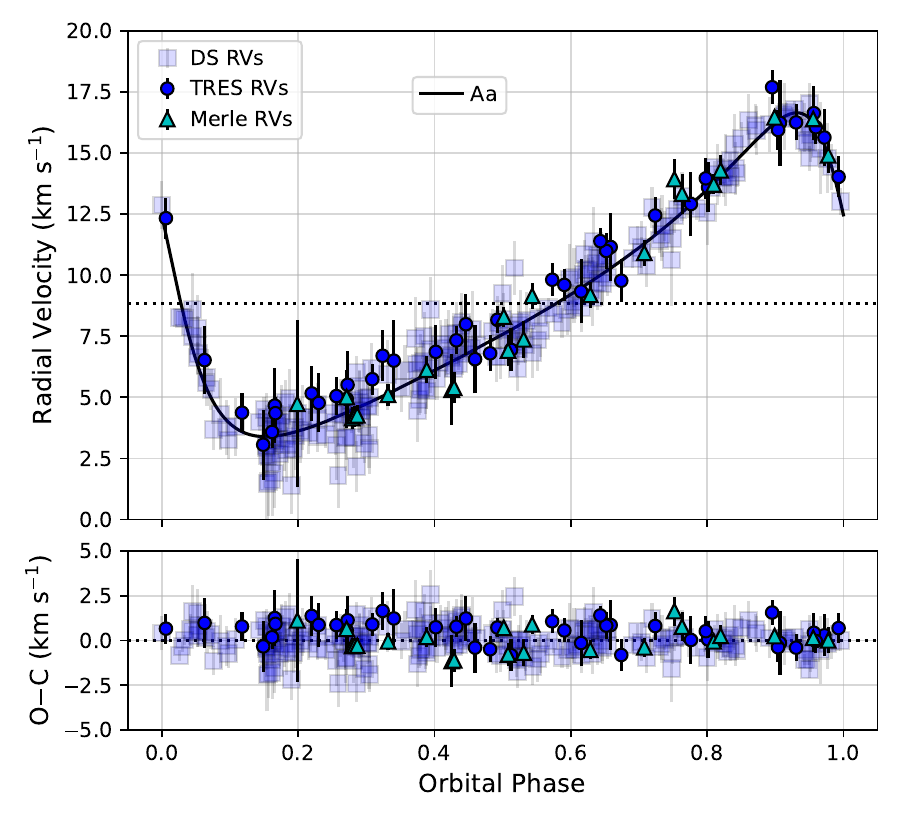}
\figcaption{RV measurements for the \hda, after removing the
motion in the outer orbit. Phase 0.0 corresponds to periastron.
The solid curve is our model, and
the dotted line marks the center-of-mass velocity. Velocity
residuals are shown at the bottom.\label{fig:RVA}}
\end{figure}

\begin{figure}
\includegraphics[width=\columnwidth]{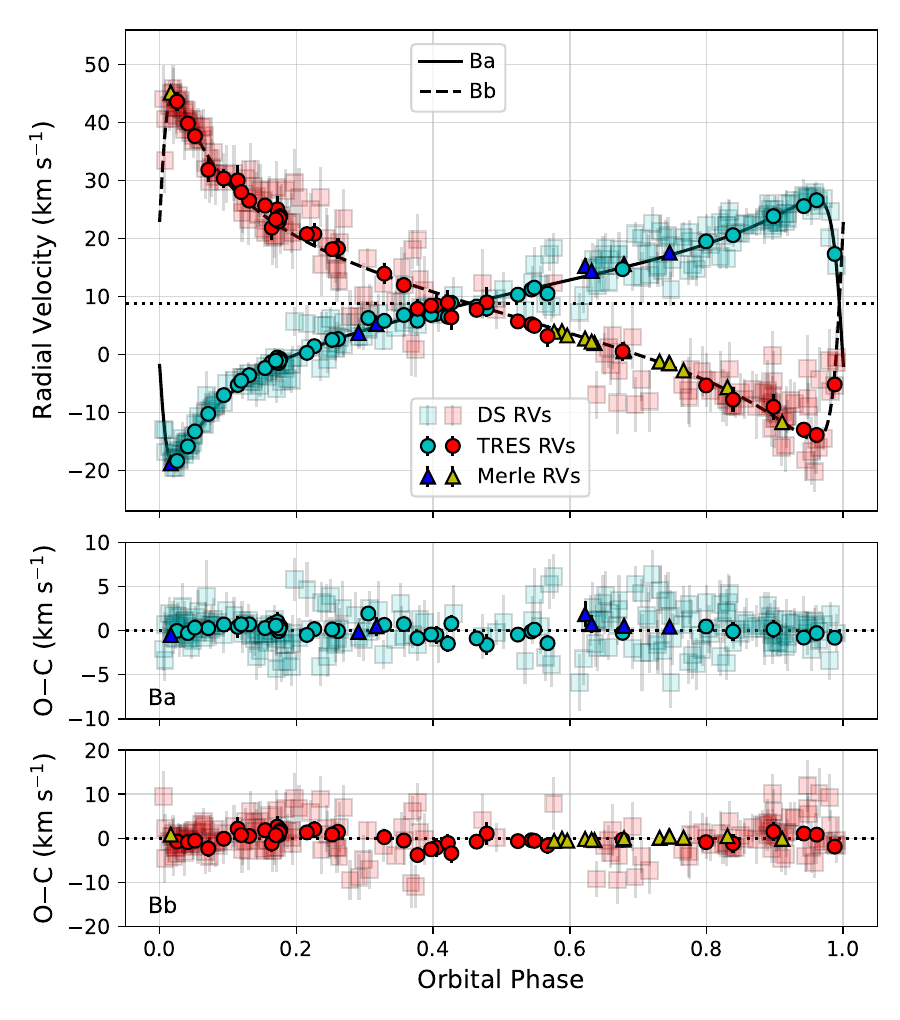}
\figcaption{Similar to Figure~\ref{fig:RVA}, for \hdb.\label{fig:RVB}}
\end{figure}

\begin{figure}
    \includegraphics[width=0.47\textwidth]{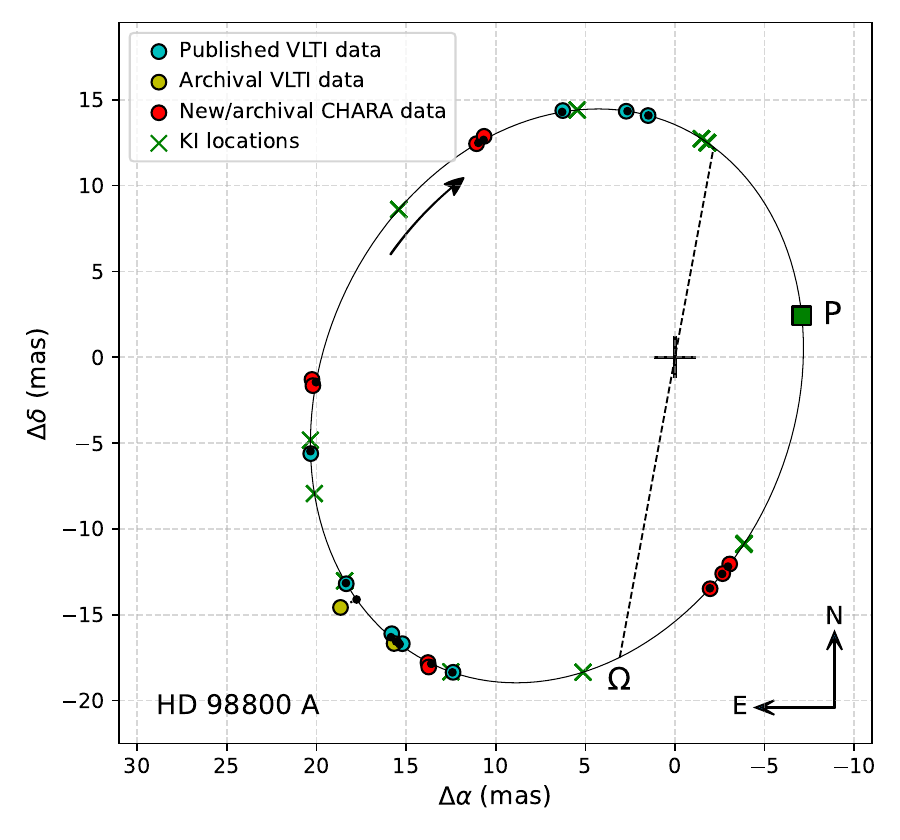}
    \figcaption{Our model for the astrometric orbit of \hda, 
    compared with published, new, and archival measurements of the relative
    position from CHARA and the VLTI. The uncertainties are typically smaller than the symbol size. The dotted line through the origin represents the
    line of nodes, with the ascending node labeled ``$\Omega$''.
    Periastron is marked with a square symbol. The KI visibility
    measurements, which cannot be represented in this plot, are
    indicated with cross symbols at their predicted locations.
    \label{fig:Asky}}
\end{figure}

\begin{figure}
    \includegraphics[width=0.47\textwidth]{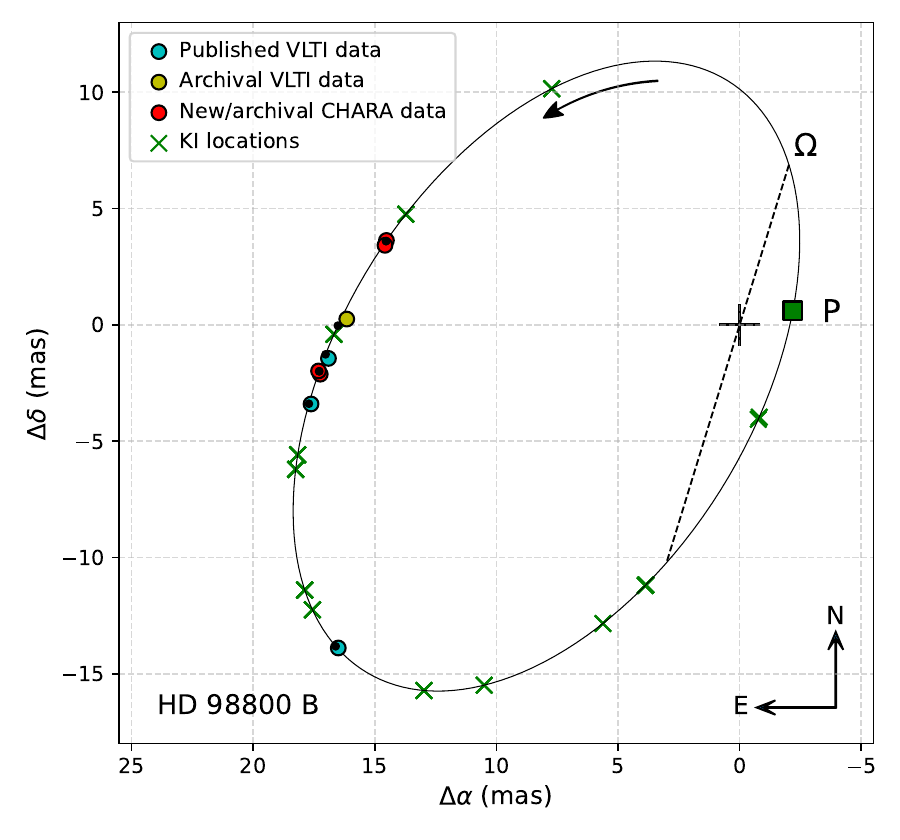}
    \figcaption{Similar to Figure~\ref{fig:Asky}, for \hdb.
    \label{fig:Bsky}}
\end{figure}

Figure~\ref{fig:RVA} and Figure~\ref{fig:RVB} display our spectroscopic models for the inner orbits, along with our RVs and those of \cite{Merle:2024}. The astrometric orbits for \hda\ and B are shown in Figure~\ref{fig:Asky} and Figure~\ref{fig:Bsky}, with the CHARA and VLTI measurements from Section~\ref{sec:longbaseline}. The predicted locations of the KI measurements on the orbits are indicated with crosses, as the visibilities themselves cannot be represented. The fits to the $V^2$ measurements of A and B on each night are shown separately in {Appendix~B}. 
The KI data provide a valuable complement to the phase coverage of the CHARA and VLTI measurements, particularly for \hdb.

The individual masses we derive for the components are presented
in Table~\ref{tab:mass}. The values for Aa and Ab are
similar to those of \cite{Zuniga:2026}, but significantly more
precise. Our masses for stars
Ba and Bb are closer to those of \cite{Zuniga:2026} than to those of
\cite{Boden:2005}, and again, significantly more precise than either
of them.
Our analysis indicates that the total mass of \hdb\ is larger than
that of the A binary. Nevertheless, \hda\ is the brighter one at
optical wavelengths. At near-infrared wavelengths ($H$ and $K$ bands),
their brightness is similar, as we illustrate below in another figure.

The orbital parallax we derive for the system, corresponding to a distance of about 46\,pc, is compared in Table~\ref{tab:parallax} against
other available estimates from the Hipparcos and Gaia missions.

\setlength{\tabcolsep}{3pt}
\begin{deluxetable}{lcc}
\tablewidth{0.95\columnwidth}
  \tablecaption{Distance, Relative Inclination Angles, and Mass-Related Properties of \hd\ \label{tab:mass}}
  \tablehead{
    \colhead{Parameter} &
    \colhead{This work} &
    \colhead{Previous results}
  }
  \startdata
  $\pi_{\rm orb}$ (mas)                     &  $21.80 \pm 0.16$\phn  & \phm{\tablenotemark{a}}$21.90 \pm 0.35$\tablenotemark{a}\phn \\
  Distance (pc)                             &  $45.86 \pm 0.34$\phn  & \phm{\tablenotemark{a}}$45.66 \pm 0.71$\tablenotemark{a}\phn \\ [+0.5ex]
  \hline \\ [-1.5ex]
  \multicolumn{3}{c}{Mutual Inclination Angles} \\ [0.5ex]
  \hline \\ [-1.5ex]
  $\varphi_{\rm A,AB}$ (degree)                &  $49.21 \pm 0.15$      & \phm{\tablenotemark{b}}$49.2 \pm 0.1$\tablenotemark{b}\phn \\
  $\varphi_{\rm B,AB}$ (degree)                &  $146.63 \pm 0.18$\phn & \phm{\tablenotemark{b}}$146.8 \pm 0.5$\tablenotemark{b}\phn\phn \\
  $\varphi_{\rm A,B}$ (degree)                 &  $157.15 \pm 0.27$\phn & \phm{\tablenotemark{b}}$157.3 \pm 0.5$\tablenotemark{b}\phn\phn \\ [+0.5ex]
  \hline \\ [-1.5ex]
  \multicolumn{3}{c}{Mass-Related Properties} \\ [0.5ex]
  \hline \\ [-1.5ex]  
  $M_{\rm Aa}$ ($M_{\sun}$)                 &  $0.975 \pm 0.025$        & \phm{\tablenotemark{b}}$0.956 \pm 0.059$\tablenotemark{a} \\
  $M_{\rm Ab}$ ($M_{\sun}$)                 &  $0.2921 \pm 0.0051$      & \phm{\tablenotemark{b}}$0.298 \pm 0.013$\tablenotemark{a} \\
  $M_{\rm Ba}$ ($M_{\sun}$)                 &  $0.823 \pm 0.012$        & \phm{\tablenotemark{b}}$0.760 \pm 0.020$\tablenotemark{a} \\
  $M_{\rm Bb}$ ($M_{\sun}$)                 &  $0.6303 \pm 0.0075$      & \phm{\tablenotemark{b}}$0.612 \pm 0.013$\tablenotemark{a} \\
  $M_{\rm A}$ ($M_{\sun}$)                  &  $1.267 \pm 0.029$        & \phm{\tablenotemark{b}}$1.280 \pm 0.085$\tablenotemark{c} \\
  $M_{\rm B}$ ($M_{\sun}$)                  &  $1.453 \pm 0.019$        & \phm{\tablenotemark{b}}$1.315 \pm 0.086$\tablenotemark{c} \\
  $q_{\rm A} \equiv M_{\rm Ab}/M_{\rm Aa}$  &  $0.2997 \pm 0.0045$      & \nodata \\
  $q_{\rm B} \equiv M_{\rm Bb}/M_{\rm Ba}$  &  $0.7661 \pm 0.0059$      & \nodata \\
  $q_{\rm AB} \equiv M_{\rm B}/M_{\rm A}$   &  $1.147 \pm 0.016$        & \nodata 
  \enddata
  \tablecomments{The values listed from this work were calculated directly from the MCMC chains
  for the orbital elements involved, and correspond to the median of the posterior
  distributions. The uncertainties represent the 68.3\% credible intervals.}
  \tablenotetext{a}{\cite{Zuniga:2026}, Table~1.}
  \tablenotetext{b}{\cite{Zuniga:2021}, Table~5.}
  \tablenotetext{c}{\cite{Zuniga:2026}, Table~2, Solution~II.}
\end{deluxetable}
\setlength{\tabcolsep}{6pt}

\setlength{\tabcolsep}{12pt}
\begin{deluxetable}{lc}
\tablecaption{Parallax Estimates for \hd\ \label{tab:parallax}}
\tablehead{
\colhead{Source} &
\colhead{Value (mas)}
}
\startdata
Hipparcos 1997         &  $21.43 \pm 2.86$\phn \\
Hipparcos 2007         &  $22.27 \pm 2.31$\phn \\
Gaia DR1        &  \nodata \\
Gaia DR2        &  \nodata \\
Gaia DR3, \hda\ &  $23.8 \pm 1.5$\phn \\
Gaia DR3, \hdb\ &  $20.1 \pm 1.1$\phn \\
This work       &  $21.80 \pm 0.16$\phn 
\enddata

\tablecomments{\cite{ElBadry:2025} has shown that formal uncertainties
for Gaia~DR3 parallaxes can be significantly underestimated when the
astrometric fits are poor, i.e., those with values of the RUWE
parameter \citep{Lindegren:2018} far in excess of unity. For \hd, the RUWE
values are 6.669 and 9.354 for A and B, respectively, leading to error inflation
factors of about 3.9 over the formal estimates. We have applied this
factor to obtain the uncertainties listed in this table.}
\end{deluxetable}


\section{Discussion}
\label{sec:discussion}

The considerably augmented spectroscopic and astrometric datasets,
relative to previous analyses, have allowed significant improvements
in the elements of both inner orbits of \hd. This, in turn, has strengthened
the constraint on the outer orbit, through a formulation of the model
that removes unnecessary free parameters that can be expressed in
terms of others. The absolute masses of all four stars are now known
to better than 2.6\% (Table~\ref{tab:mass}). This represents about a factor of two improvement compared to the most recent determinations by \cite{Zuniga:2026}, particularly for star Aa, whose mass was previously only known to about 6\%.
Other properties we have updated and improved on the basis of our 
global solution include the mutual inclination angles $\varphi$ between the
orbits, following the usual expression
\begin{equation}
\cos \varphi_{\rm A,B} = \cos i_{\rm A} \cos i_{\rm B} +
\sin i_{\rm A} \sin i_{\rm A}
\cos(\Omega_{\rm A} - \Omega_{\rm B}),
\end{equation}
and similar ones for the orientation of the inner orbits relative
to the outer one.
These angles are listed in the middle section of the same table. 

The circumbinary dust disk around \hdb\ was determined by
\cite{Kennedy:2019} to have an inclination angle relative to the
line of sight of either $26\fdg0 \pm 0\fdg4$ or $154\fdg0 \pm 0\fdg4$,
where the ambiguity comes from the unknown direction of rotation
of the disk as projected on the sky (clockwise or anti-clockwise).
The position angle of the disk was determined to be $15\fdg6 \pm 0\fdg8$,
corresponding to $\Omega_{\rm disk} = 195\fdg6 \pm 0\fdg8$ when using
the classical convention for the ascending node adopted in our work.
Based on simulations of the disk dynamics that disfavor intermediate
orientations, \cite{Kennedy:2019} concluded that the disk was much
more likely to be in a polar orientation (favoring the line-of-sight inclination of 26\fdg0). With these parameters,
and their own orbital solution for \hdb, \cite{Zuniga:2021} 
estimated the inclination between the disk and the binary orbit to be
$89\arcdeg \pm 1\arcdeg$, consistent with being perfectly polar.
The revised elements of the B orbit from the present work give a
similar angle of $\varphi_{\rm disk,B} = 88\fdg74 \pm 0\fdg42$.

\subsection{Dynamical Evolution}
\label{sec:dynamics}

\cite{Zuniga:2021} used $N$-body simulations to study the dynamical stability of the \hd\ quadruple star system over the short term and the long term. They explored integration times between 1000 and $10^6$\,yr, and concluded that the system is stable over many thousands of AB orbits. Here we are interested in quantifying changes in the orbital elements of \hda\ and B, motivated in part by the possible systematic errors and excess scatter mentioned earlier in the KI observations of the B component, and to a lesser extent also in \hda. In particular we wish to investigate the behavior of the A and B  orbits on either side of the recent 2021 periastron passage in the outer orbit, and any influence the circumbinary disk may have on the B orbit. We also revisit the longer-term evolution up to the 10\,Myr age {of} the system with a more complete model of four stars.

We converted the elements of Table~\ref{tab:mcmc} into barycentric Cartesian
initial conditions for the four stars at J2000.0 and integrated the four-body
problem with the \texttt{IAS15} integrator \citep{Rein:2015} in the
\texttt{REBOUND} package \citep{Rein:2012}. This setup conserved the total energy to
better than one part in $10^{10}$ over the longest integration used here.  The
model reproduces all six velocity semiamplitudes to better than
$0.01$~km~s$^{-1}$, the three node position angles to $0\fdg01$, and the mutual
inclinations of Table~\ref{tab:mass} to $0\fdg02$. Independently of those, it
matches the interferometric positions of Table~\ref{tab:interf} with an rms of
$0.5$~mas and the CfA velocities with an rms of $1.6$~km~s$^{-1}$, and places
HD~98800~B in front of A at the present epoch, as the eclipse requires.  The
two inner centers of mass make their closest approach at 2021.44, separated by
$29.33$~au, in agreement with the fitted $T_{\rm peri,AB} = 2021.45 \pm 0.40$.

\begin{figure}
\epsscale{1.15}
\includegraphics[width=\columnwidth]{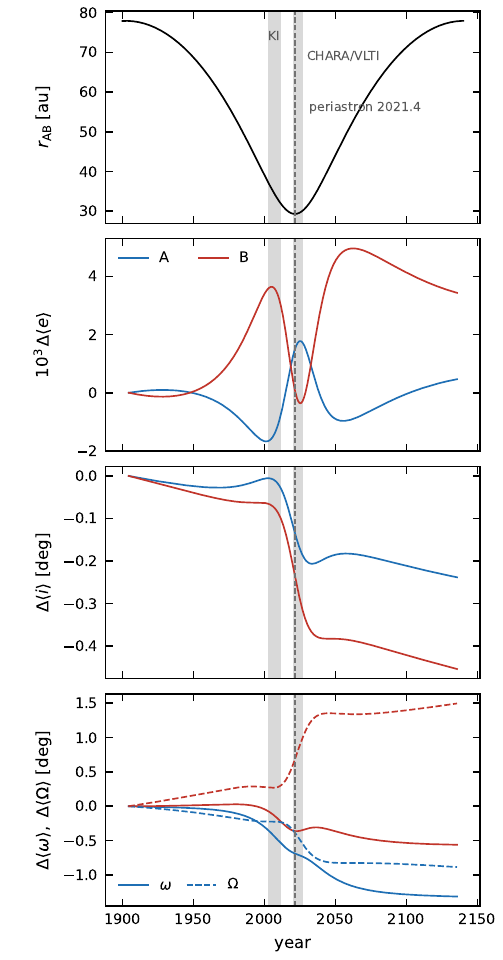}
\figcaption{The 2021 periastron passage of the outer orbit and its effect on the
inner binaries; blue is \hda\ and red \hdb.  \emph{From the top:} the
separation of the two inner centers of mass; and the changes in the
period-averaged eccentricity, inclination, argument of periastron (solid) and
node (dashed) of each inner binary, all measured relative to their 1900 values.
The dashed vertical line marks periastron at 2021.44 and the shaded bands the KI
and CHARA/VLTI observations.  Much of the excursion in $e$ is reversible; the
changes in $i$, $\omega$ and $\Omega$ are largely permanent, but reach only
$1\fdg5$ at most.  The semimajor axes, not plotted, are conserved to better than
two parts in $10^{7}$.\label{fig:dynencounter}}
\end{figure}

The encounter is slow compared with the inner orbits: the characteristic
encounter time $(q_{\rm AB}^{3}/GM_{\rm tot})^{1/2}$ is about 15~yr, some
18--21 times the 0.72 and 0.86~yr periods of the A and B binaries, so the
passage is adiabatic for the inner pairs and large permanent changes are not
expected.  Integrating from 1900 to 2140 --- from one apoapsis of the outer
orbit, through periastron, to the next --- and comparing period-averaged
osculating elements in the two apoapsis windows gives the permanent change
(Figure~\ref{fig:dynencounter}).  We find that the semimajor axes are conserved to $\Delta
a/a = 6.5\times10^{-8}$ (A) and $1.7\times10^{-7}$ (B), so the inner periods
are untouched.  The eccentricities change by $+5.5\times10^{-4}$ and
$+3.3\times10^{-3}$, and the orientation angles by at most $1\fdg5$.  Monte
Carlo realizations drawn from the uncertainties of Table~\ref{tab:mcmc} give
fractional spreads of a few per cent on these numbers.  The B binary, being
both wider and much more eccentric, is perturbed more strongly, by a factor
$6.1^{+1.2}_{-1.1}$ in the permanent eccentricity change.

To assess how constraining the KI visibilities are, we used the measurements on each night to determine a mean relative position and its uncertainty as an error ellipse.\footnote{Note that our global solution does not use nightly relative positions, but instead incorporates the visibilities directly.} Those ellipses prove to be very elongated, because the KI has a single baseline whose position angle barely rotates. The minor axes lie at $43\arcdeg$--$47\arcdeg$ on every night for both inner binaries, and measure 0.16\,mas for A and 0.20\,mas for B in the median. Their major axes are ten times larger: about 1.6\,mas for A and 2.1\,mas for B. The more recent CHARA/VLTI measurements have median uncertainties of about 0.09 and 0.3\,mas for A and B, in each coordinate (Table~\ref{tab:interf}).

Against those uncertainties the four-body signal is small.  We sampled the
relative position of each inner binary from the integration at the real epochs
of observation.  Fitting a single Keplerian to the whole 2003--2026 span, as
our global solution does, leaves systematic residuals of only $8\,\mu$as for A
and $16\,\mu$as at the modern epochs for B --- more than an order of magnitude
below the measurement uncertainties quoted above.  The dynamics {do} predict a
somewhat larger effect for B than for A, less than the $6.1$ factor in the elements,
since a fit absorbs whatever a shift of the elements can mimic; but it is far
too small to be the explanation for the large residuals in the KI data.

Table~\ref{tab:dynlimits} states the same conclusion element by element.  For
each element we computed the on-sky displacement produced by a unit change and
inverted it to find the change a given astrometric precision could detect.
Measured against the CHARA/VLTI uncertainties actually achieved --- $0.09$~mas for A and
$0.3$~mas for B per coordinate --- every predicted change lies below the
detection threshold, the largest ratio being $0.8$, for $\Delta\omega$ of A.
Changes large enough to matter would in any case have been conspicuous in
36~yr of radial velocity monitoring.  The perturbations also vary smoothly, on the
timescale of the inner periods and of the 238-yr outer orbit, so no dynamical
mechanism can produce a discrepancy confined to a single night, such as the one
on MJD~53101, while leaving its neighbors untouched.  We conclude that the
systematic errors in the visibilities are not of dynamical origin.

Two related statements should be kept apart, however.  The perturbations do not
degrade the global solution: a constant-element Keplerian reproduces the
four-body motion at the observed epochs to well within the measurement
uncertainties, so no systematic residual pattern is introduced, and the fit is
not biased in the sense that matters for the visibilities.  But the elements it
returns are averages over the interval the data cover, and across that interval
the perturbations move them by an amount that is not negligible against their
formal errors: between one and four times the uncertainties quoted in
Table~\ref{tab:mcmc}, most sharply for $e_{\rm A}$ ($4.4\sigma$) and
$\omega_{\rm A}$ ($3.9\sigma$), and least for $i_{\rm A}$ ($1.0\sigma$).  The
elements of Table~\ref{tab:mcmc} are therefore best read as epoch-averaged
rather than instantaneous osculating values.

The velocities provide an independent check, since they measure times rather
than positions.  Tracking successive periastron passages across the 36\,yr
spectroscopic baseline, the perturbations make those times wander about a
linear ephemeris with an rms of $0.035$~d for A and $0.114$~d for B.  For A
this is well inside the quoted uncertainty on $T_{\rm peri,A}$; for B it is
about twice the formal $\sigma(T_{\rm peri,B}) = 0.061$~d, so the accuracy of
that epoch, as opposed to its precision, may be set by unmodeled dynamics.
Even so, the effect is invisible in the velocities themselves, corresponding at
the steepest part of the B velocity curve to at most $0.7$~km~s$^{-1}$.

Over longer times the system is stable but not static.  Whether an inner binary
undergoes von~Zeipel--Lidov--Kozai (ZLK) oscillations is set not by its inclination
$\varphi$ alone but by $\Theta = \sqrt{1-e^{2}}\cos\varphi$, which the quadrupole
interaction conserves, with libration possible when $\Theta^{2} < 3/5$; the
familiar window $39\fdg2 < \varphi < 140\fdg8$ is the circular-orbit case of that
condition.  Although $\varphi_{\rm B,AB}$ lies outside that window, $\Theta_{\rm
A} = +0.57$ and $\Theta_{\rm B} = -0.52$ are both well inside the condition
itself, so both binaries are ZLK-active.  A $10^{7}$~yr integration, spanning
the age of the parent TW\,Hya association, bears this out: binary A cycles between $e_{\rm A} = 0.27$
and $0.61$ with power at $4.3$ and $5.2\times10^{4}$~yr, and B cycles between $0.44$
and $0.81$ every $3.5\times10^{4}$~yr, while all three semimajor axes hold
constant to a few parts in $10^{4}$.  Measured from the mutual node, the
argument of periastron of A librates about $270\arcdeg$ with a half-amplitude
of $23\arcdeg$ and never circulates; that of B circulates steadily every
$7.0\times10^{4}$~yr.  The eccentricities we measure are therefore phases of an
ongoing cycle, not primordial values.  The outer orbit is steady as well: $i_{\rm
AB}$ oscillates only between $84\fdg5$ and $92\fdg0$, and the smallest
sky-projected separation reached by A from the center of B never exceeds
$3.6$~au, always within the dust disk.  A close projected alignment therefore
occurs at essentially every periastron passage, once every $\sim\!238$~yr.
Which component is in front alternates as the outer apsidal line precesses,
with a period of $4.5\times10^{6}$~yr, so A is the occulted one in about half
of them.  The event we are now witnessing is a recurring feature of the
architecture rather than a coincidence of the present epoch.

That leaves the disk itself, which B has and A does not.  We modeled it in the
polar orientation determined above, as a ring exerting a secular torque on the
Ba--Bb pair, and estimated the back-reaction two ways: from the quadrupole
precession rate, and from angular-momentum reciprocity, using the disk's own
precession frequency measured from test-particle integrations.  The two agree
to within a factor of three.  With the masses derived by \citet{Kennedy:2019}
--- lower bounds of $0.33\,M_\oplus$ in dust and $0.28\,M_\oplus$ in gas,
between 2.5 and 4.6~au --- the disk reorients the binary by $0.1$--$0.4\,\mu$as
over the whole 2003--2026 astrometric baseline.

Since the effect is linear in the disk mass, this inverts into a requirement.
To move the B orbit by as much as the best astrometric uncertainty of 0.17\,mas  actually achieved for B, the disk would have to weigh $0.9$--$2.4\,M_{\rm
Jup}$, some $500$--$1200$ times the measured lower bound and a fraction of a per
cent of the mass of the B binary, confined between 2.5 and 4.6~au.  The
millimeter emission is optically thick and the CO-to-gas conversion uncertain
\citep{Ribas:2026}, so the true mass is certainly higher than that lower bound,
but it would have to be higher by a factor of order $10^{3}$ before the disk
had any measurable effect on the orbit.  The disk is not the source of the
larger scatter in the B visibilities.

The binary controls the orientation of the disk.  Test particles displaced from
the polar state librate about it \citep{Farago:2010, Martin:2017} with periods
of $52$~yr at the inner dust edge and $315$~yr at the outer, far shorter than
the $\sim\!1.7\times10^{4}$~yr on which the outer orbit can reorient the B
binary, so the disk follows the binary adiabatically and stays polar.  Those
same integrations bound the disk radially: coplanar test particles do not
reliably survive inside $4.2$~au, in good agreement with the criterion of
\citet{Holman:1999}, whereas polar ones survive down to $2.6$~au, and material
around B is stripped by A outside $7.33$~au.  The observed dust, between 2.5
and 4.6~au, fills that polar annulus almost exactly --- the size of the central
cavity is itself an indication of the polar geometry, and suggests it is set by
dynamical clearing.

\begin{deluxetable*}{lcccc}
\tablecaption{Predicted and Detectable Changes in the Inner Orbits of \hd
\label{tab:dynlimits}}
\tablehead{
  \colhead{} & \multicolumn{2}{c}{\hda} & \multicolumn{2}{c}{\hdb} \\ [0.5ex]
  \cline{2-3}\cline{4-5} \\ [-1ex]
  \colhead{Element} & \colhead{Predicted} & \colhead{Detectable} &
  \colhead{Predicted} & \colhead{Detectable}}
\startdata
$\Delta e$              & $+0.00315 \pm 0.00018$ & $0.0053$ & $-0.00385 \pm 0.00023$ & $0.016$ \\
$\Delta i$ (deg)        & $-0.157 \pm 0.017$     & $0.58$   & $-0.208 \pm 0.024$     & $0.93$  \\
$\Delta \Omega$ (deg)   & $-0.257 \pm 0.047$     & $0.40$   & $+0.602 \pm 0.090$     & $1.46$  \\
$\Delta \omega$ (deg)   & $-0.286 \pm 0.029$     & $0.36$   & $-0.227 \pm 0.027$     & $0.84$  
\enddata
\tablecomments{``Predicted'' is the change in the period-averaged osculating
element between the KI epochs (2003--2010) and the CHARA/VLTI epochs
(2021--2026), from the four-body integration. The uncertainties are the
$16$--$84$ percentile range over $200$ Monte Carlo realizations of the elements
of Table~\ref{tab:mcmc}.  ``Detectable'' is the change that would displace the
relative orbit on the sky by the actual median CHARA/VLTI position uncertainty of
Table~\ref{tab:interf}, $0.09$~mas for A and $0.3$~mas for B per coordinate,
taken here as a total displacement and so multiplied by $\sqrt{2}$.  Every
predicted change lies below the corresponding threshold, the largest ratio
being $0.8$ for $\Delta\omega$ of A.  The semimajor axes change by less than two parts in
$10^{7}$ and are omitted.}
\end{deluxetable*}

\subsection{Comparison with Stellar Evolution Models}
\label{sec:stellarevolution}

\begin{figure}
\includegraphics[width=\columnwidth]{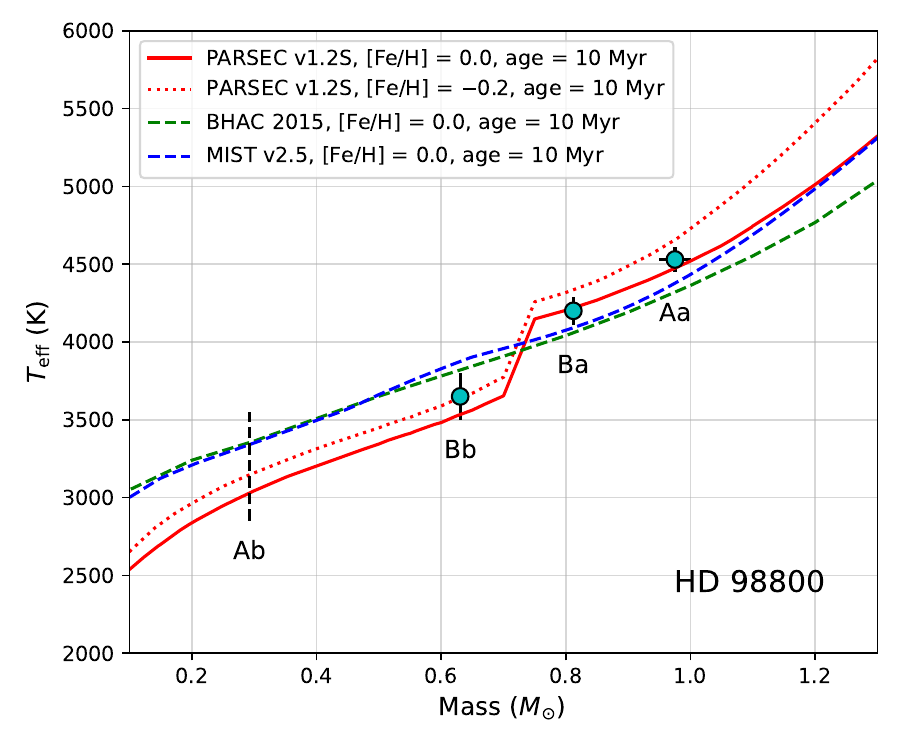}
\figcaption{Empirical mass and temperature estimates compared
against stellar evolution models. The masses are from this work, and
the effective temperatures are from the work of \cite{Laskar:2009}.
Star Ab lacks a temperature determination because it has not been detected
spectroscopically. We indicate its mass on the plot with a vertical dashed line.
Isochrones from the PARSEC~v1.2S series
\citep{Chen:2014} are shown for an age of 10\,Myr and two different
compositions: solar, and ${\rm [Fe/H]} = -0.20$. Two additional models
for solar metallicity are from the MIST~v2.5 series \citep{Choi:2016}
and \cite{Baraffe:2015}. \label{fig:massteff}}
\end{figure}

The spectroscopic properties of the three visible components of \hd\
were studied most recently by \cite{Laskar:2009}, who reported effective temperatures
for Aa, Ba, and Bb of $4530 \pm 80$\,K, $4200 \pm 90$\,K, and $\sim$3650\,K,
respectively, where the uncertainties are statistical only.
Their estimate for Bb was hindered by the limited number
of unblended spectral lines available for that component. We arbitrarily
adopted for that star a conservative uncertainty of 150\,K. They also reported
the metallicity of the system to be slightly subsolar: ${\rm [Fe/H]} = -0.20 \pm 0.10$.
In Figure~\ref{fig:massteff}, we compare our mass measurements and the
above temperatures against several models of stellar evolution, for the
consensus age of 10~Myr adopted for the TW~Hya association. The agreement
is generally good, with a slight preference for solar composition
when using the PARSEC~v1.2S models of \cite{Chen:2014}. These models display
an obvious discontinuity (see the figure) that is caused by the adoption of a modified
temperature-opacity relation for stars less massive than about 0.7\,$M_{\odot}$.
This change was introduced by the modelers to improve the match between the predicted radii and the observed radii of low-mass stars. The other two models shown in the figure also represent the observations reasonably well.

\begin{figure}
\includegraphics[width=\columnwidth]{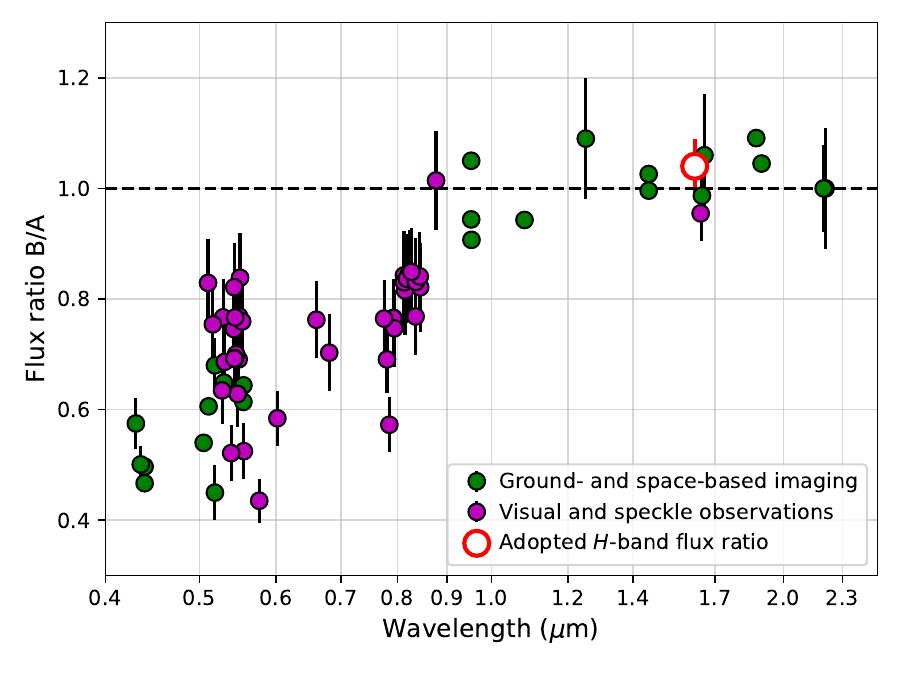}
\figcaption{Ground-based and space-based measurements of the flux
  ratio between \hda\ and B {from \cite{Soderblom:1998},
 \cite{Low:1999}, \cite{Prato:2001}, Hipparcos, Tycho-2, and the visual and speckle observers.}
  The value adopted for the $H$ band is $1.03 \pm 0.05$, {calculated from a simple
  average of measurements redward of 1.2\,$\mu$m, and assigned a conservative uncertainty}.\label{fig:fluxrat}}
\end{figure}

\begin{figure}
\includegraphics[width=\columnwidth]{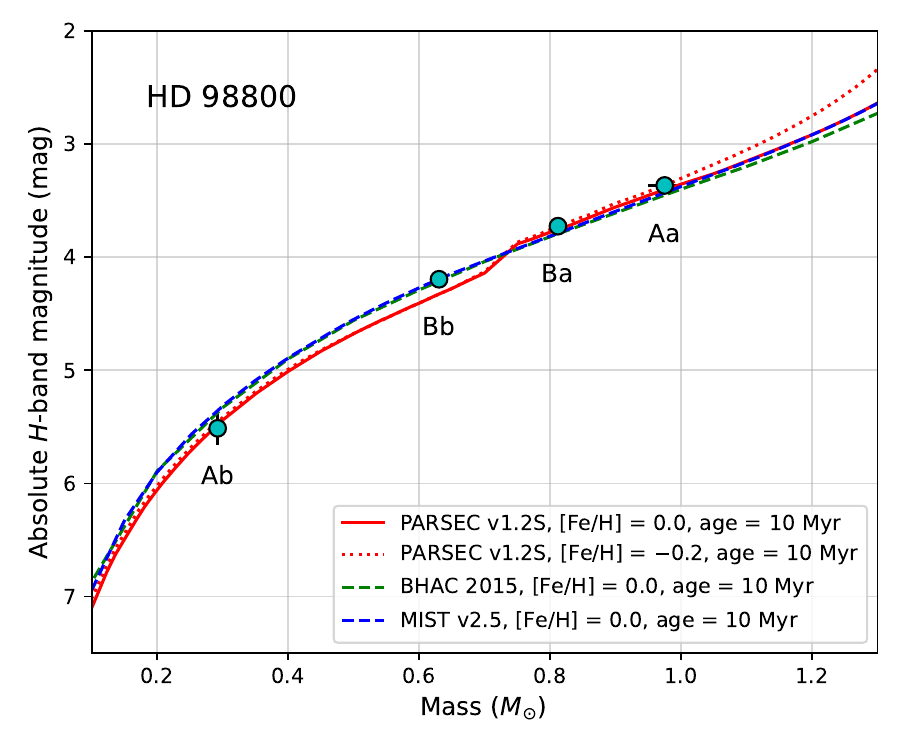}
\figcaption{Absolute $H$-band magnitudes for \hd\ versus mass. {In most
  cases the uncertainties are smaller than the point sizes.} The model
  isochrones are the same as shown in Figure~\ref{fig:massteff}.
\label{fig:massH}}
\end{figure}

An additional comparison with theory can be made
using near-infrared brightness measurements for the four components.
The interferometric observations from this work and from \cite{Zuniga:2021, Zuniga:2026} provide the $H$-band flux
ratios between the primary and secondary within each binary, averaging
$f_{\rm Ab}/f_{\rm Aa} = 0.14 \pm 0.02$ and 
$f_{\rm Bb}/f_{\rm Ba} = 0.65 \pm 0.03$.
The brightness difference in $H$ between \hda\ and \hdb\ has been measured {in the optical and near-infrared} at
multiple wavelengths both from the ground and from space. {These measurements do not resolve the individual components within each binary.} We adopted a mean
for the $H$ band of $f_{\rm B}/f_{\rm A} = 1.03 \pm 0.05$ (see Figure~\ref{fig:fluxrat}).
Finally, with the total $H$-band
brightness from the 2MASS catalog ($5.759 \pm 0.027$, ignoring extinction),
and our orbital parallax, we derived
the following absolute magnitudes for each star: $3.36 \pm 0.05$ (Aa),
$5.51 \pm 0.15$ (Ab), $3.73 \pm 0.05$ (Ba), and $4.20 \pm 0.05$ (Bb).
Figure~\ref{fig:massH} shows these measurements as a function of mass,
compared with the same models as in Figure~\ref{fig:massteff}. The
agreement is very good, with little difference in this case between
the PARSEC~v1.2S models for solar metallicity and for ${\rm [Fe/H]} = -0.20$.

\subsection{The {occultation} of \hda}
\label{sec:eclipseA}

\hda\ and \hdb\ will soon reach their closest separation on the plane
of the sky, as they revolve around each other with a period of $\sim$240\,yr.
The nearly edge-on orientation of the outer orbit takes
the path of A directly behind the circumbinary disk surrounding B,
producing a once-in-a-lifetime opportunity to witness this eclipse
event, and to study the structure and composition of the disk.

The expected time evolution of the brightness of the system was studied 
previously by \cite{Zuniga:2021}, using the results of their orbital
analyses of the inner and outer orbits, the knowledge of the
component masses as it stood then, and the disk properties as reported
by \cite{Kennedy:2019}, which are based on their ALMA observations. According to
the latter authors, the dust is located between 2.5 and 4.6\,au from the
center of mass of \hdb\ (for their assumed distance of 44.9\,pc, slightly
shorter than ours), while the gas extends both further out and further in.
\cite{Zuniga:2021} predicted the event would
start in 2026, finishing sometime between 2030 and 2031. Their numerical
simulations indicated that halfway through this interval the A binary will
traverse the central cavity of the disk, temporarily returning the
brightness of the system to a higher level, before reducing it again
as it passes behind the other side of the disk. Refined time windows for the various stages of the eclipse, such as ingress and egress from the dust and gas components of the disk, were presented by \cite{Zuniga:2026}.

Recent studies have contributed to our knowledge of the disk itself by
using hydrodynamical simulations to
explore the effects of varying the mass of the gas and dust components, as well as the viscosity \citep{Faruqi:2025}. Others
have investigated the effects of the illumination of the disk by stars Ba
and Bb over the course of their orbit, given the fact that the orbital
and disk planes are nearly at right angles to each other \citep{Rabago:2025}.

In the present work we have significantly improved the determination of the
three orbits, the distance to the system, and the masses of all four components.
While these refinements do not change the overall features of the event,
they do have an impact on the details, and
they provide a much more accurate understanding of the
dynamics of the system, which will be valuable as the eclipse progresses.
And as is evident from the observations, the eclipse is in fact already underway,
as anticipated by \cite{Zuniga:2021}.

\begin{figure}
\includegraphics[width=\columnwidth]{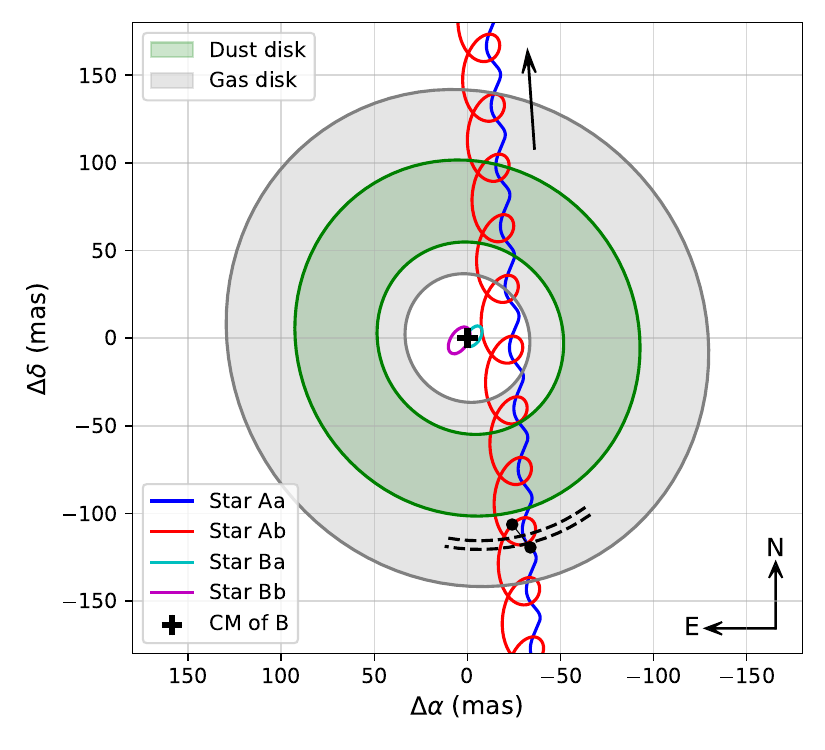}
\figcaption{Illustration of the paths of stars Aa and Ab as they approach
the circumbinary disk around \hdb\ from the south, similar to a figure presented by
\cite{Zuniga:2021}. The origin is fixed at the center of mass (CM)
of B. Near the bottom of the figure, the
dots on the trajectories represent the positions of the stars on the date
of a recent CHARA observation, on UT 2026\,Mar\,25 (see the text). {The two dashed lines mark the radial locations at which star Aa began to decline in brightness around mid April of 2026, and then became invisible in the
optical, approximately on 2026\,May\,6 (see Figure~\ref{fig:lightcurve}).}
\label{fig:approach}}
\end{figure}

Figure~\ref{fig:approach} is an updated version of a similar illustration
by the above authors, showing the expected projected paths of \hd\,Aa and Ab throughout the
event, based on the orbits derived in this paper. The origin is placed
at the center of mass of \hdb. The gas and dust disks are drawn with
their sizes, shapes, and orientations as reported by \cite{Kennedy:2019}.

\begin{figure}
\includegraphics[trim={40pt 30pt 40pt 85pt}, clip, width=\columnwidth]{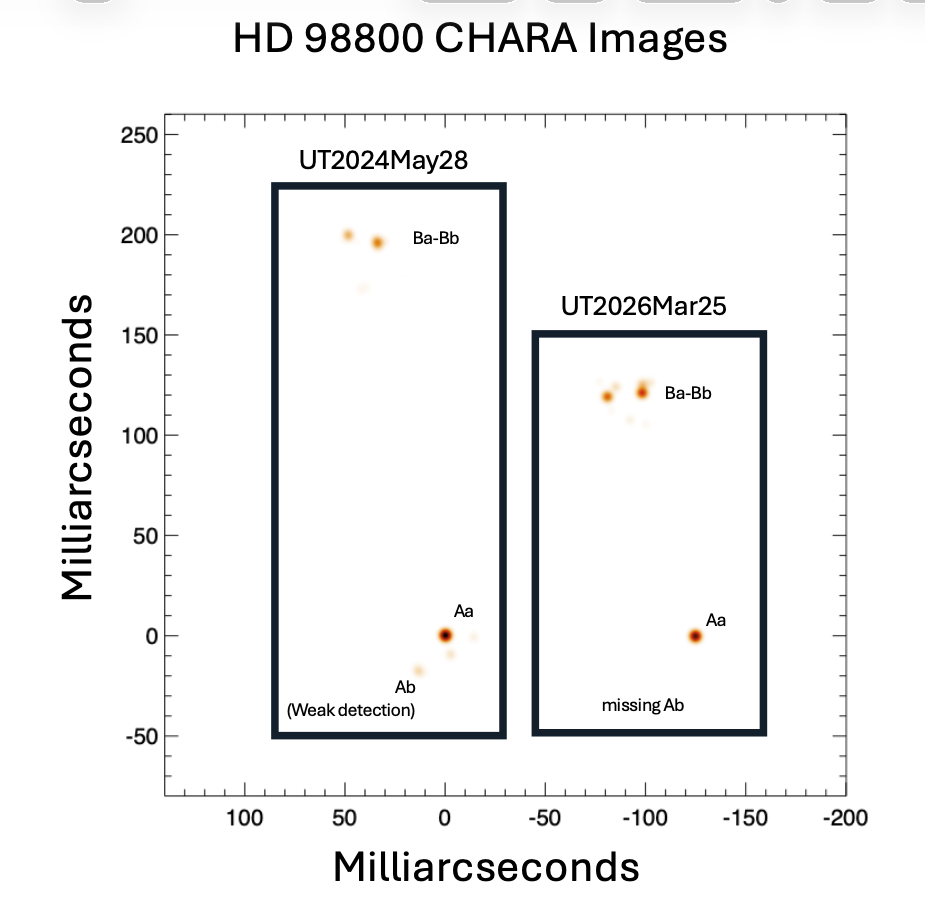}
\figcaption{Images constructed from two of our CHARA observations
obtained about two years apart, set to the same scale. On the left ($H$ band),
\hda\ is still far from being occulted by the disk, and all four
components are visible. On the right ($K$ band), star Ab is missing because
extinction from the disk has already blocked its light.
\label{fig:images}}
\end{figure}

Two of our CHARA observations have allowed us to construct images of the
system showing the A and B binaries simultaneously in the field of view.
One is from UT 2024\,May\,28, and the other from UT 2026\,Mar\,26 (see Figure~\ref{fig:images}).
In the first, taken well before \hda\ approached the disk from behind, all four stars
are visible, and while Ab is very faint in the figure, detailed modeling
recovers it at the expected position and with approximately the expected brightness.
In the second image, however, star Ab is not seen, and detailed modeling
fails to detect it as well. The positions of the components of \hda\ relative to the
disk on this date are shown by the dots near the bottom of Figure~\ref{fig:approach}, and place the
binary already well behind the gas component of the disk. Extinction caused by small, micron-size dust particles embedded within the gas are likely responsible for the disappearance of the star.

\begin{figure}
\includegraphics[width=\columnwidth]{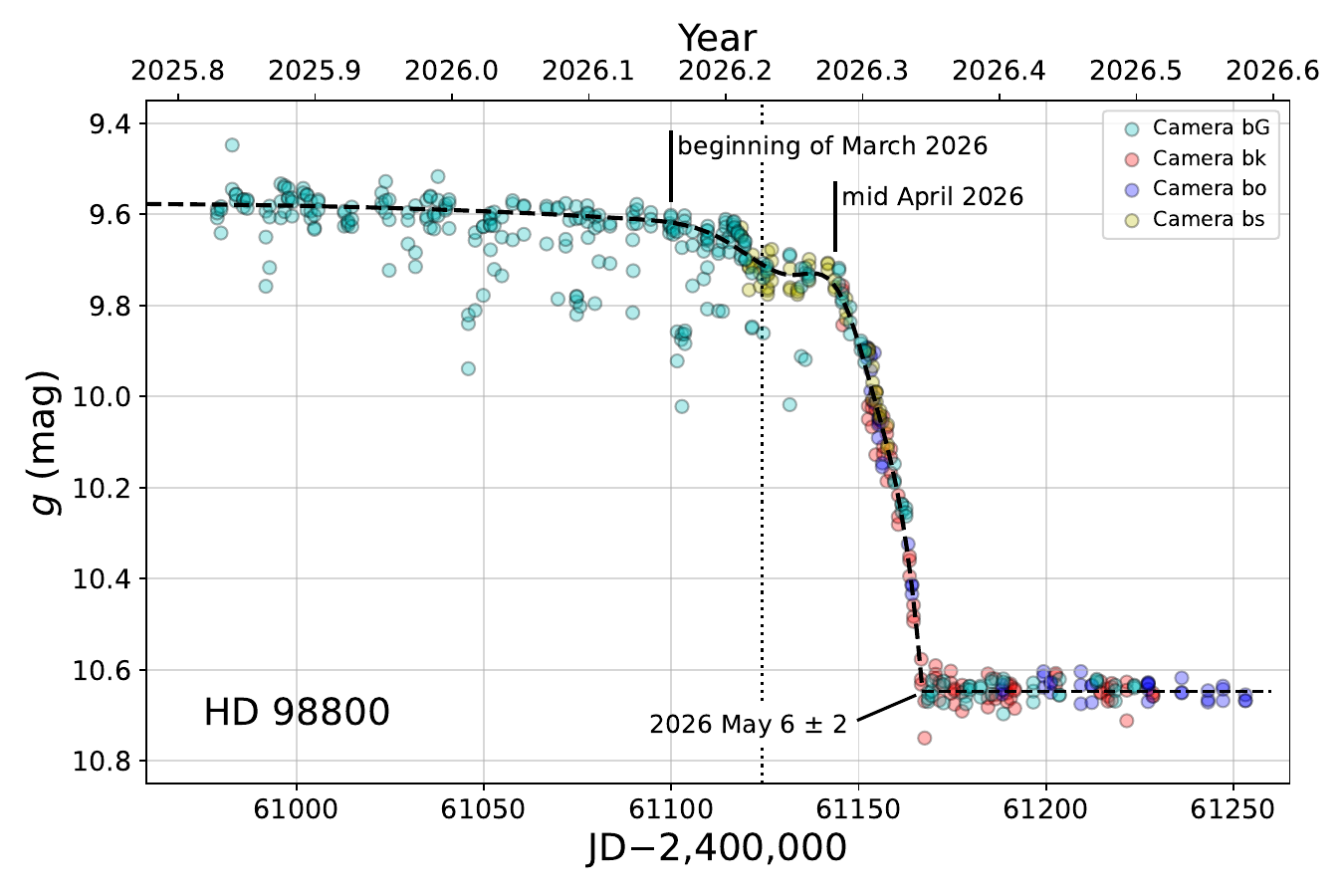}
\figcaption{Light curve of \hd\ in the Sloan $g$ band, from the ASAS-SN Sky Patrol cameras run by The Ohio State University \citep{Shappee:2014, Kochanek:2017}. A smooth curve through the data was drawn by eye. Most of the observations shown here are from the `bG' camera (cyan points), which has the smallest scatter. Observations from other cameras (bk, bo, bs) were added in the most recent part of the curve, to better map the decline. The brightness is seen to have started to drop noticeably around the beginning of 2026 March ($\sim$2026.16), and then more abruptly beginning in mid April ($\sim$2026.28). The bottom of the light curve was reached on 2026 May 6 $\pm$ 2 ($\sim$2026.34).
The vertical dotted line marks the time of the CHARA observation shown on the
right-hand side of  Figure~\ref{fig:images}, at which star Ab had already
disappeared. The steeper decline beginning shortly thereafter signaled
the occultation of star Aa.
\label{fig:lightcurve}}
\end{figure}

The ingress phase of the eclipse, which is now complete, has been recorded in the photometric monitoring of \hd\ carried out by The Ohio State University's All-Sky Automated Survey for SuperNovae \citep[ASAS-SN;][]{Shappee:2014, Kochanek:2017}. The Sloan $g$-band observations since 2025 November, obtained with several CCD cameras at different locations, are shown in Figure~\ref{fig:lightcurve}, and reveal that the brightness of the system has dropped by a full magnitude over a short period of time. There is considerable scatter in the data before the drop, and also at earlier times not shown here, which we suspect is in part instrumental or due to local observing conditions, given that it is often different for different cameras. Some of it could be real, however, as intrinsic variability from stellar activity is common among PMS stars.

The optical lightcurve provides a means of quantifying the extinction suffered by \hda\ as a function of time. We first obtained estimates of the normal apparent brightness of each component in the $g$ band, using our measured masses from Table~\ref{tab:mass}, a model isochrone from the PARSEC v1.2S series for solar metallicity, and the known distance. We then drew a smooth curve by eye through the light curve (dashed line in Figure~\ref{fig:lightcurve}), and calculated how much extinction applied to stars Aa and Ab was required at each epoch to match the total apparent magnitude as measured by ASAS-SN. We assumed for this exercise that both stars in the A binary experience the same attenuation, which is not strictly true, as will be seen below. The flat bottom of the light curve indicates the light of A binary has effectively been blocked completely, so that the overall brightness of the system is due only to stars Ba and Bb. We find that with our nominal masses for those stars, the models are able to match the measured apparent magnitude of the system at this phase very well (within better than 0.01\,mag). On the other hand, the predicted out-of-eclipse magnitude is slightly too faint, which could be due to errors in the masses or in the models. Because star Aa contributes most of the light and also has the largest mass error, we chose to adjust its mass slightly from its nominal value in order to reproduce the ASAS-SN measurements. We found that an upward adjustment of $\sim$0.5$\sigma$ in $M_{\rm Aa}$ was sufficient to provide good agreement.

\begin{figure}
\includegraphics[width=\columnwidth]{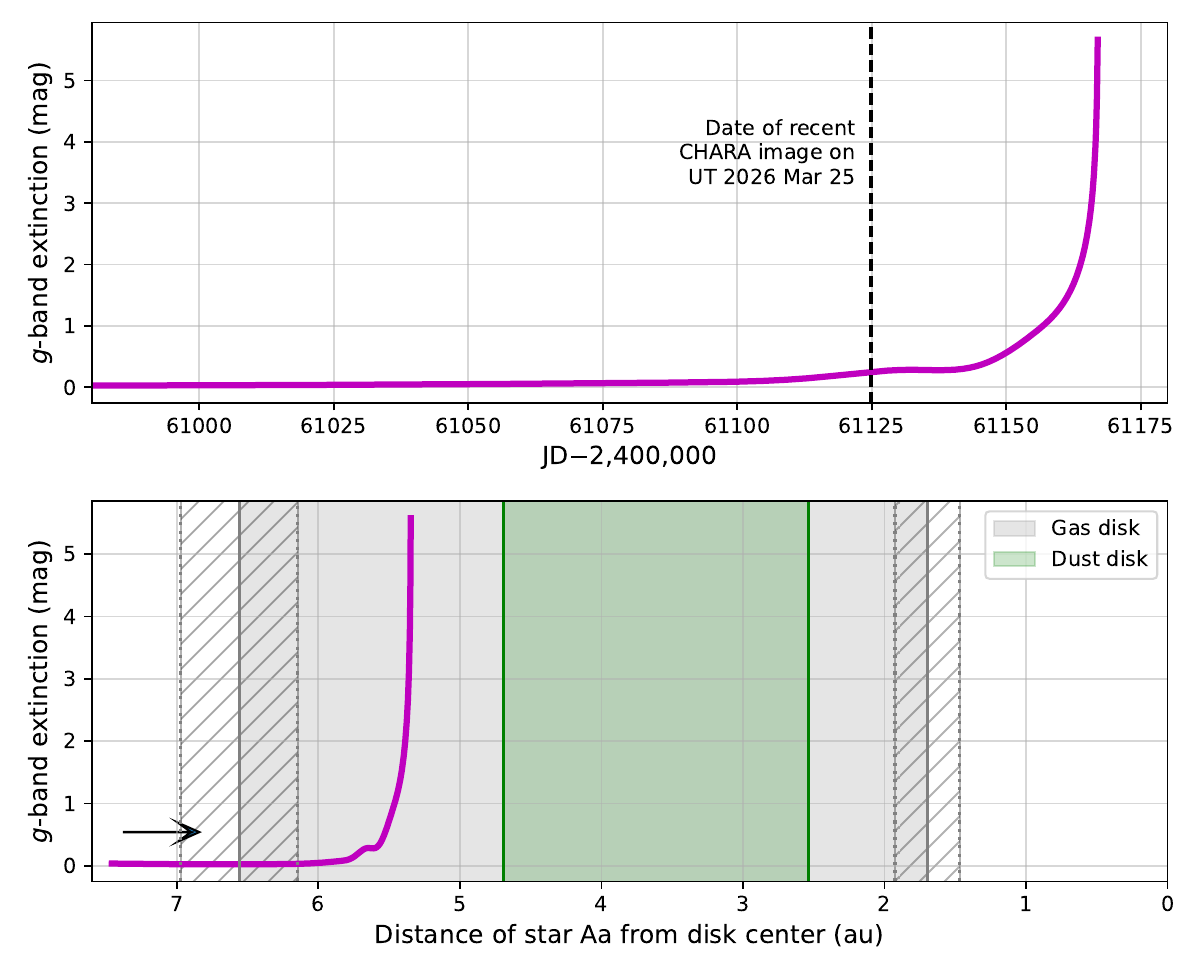}
\figcaption{The top panel shows the $g$-band extinction affecting \hda\ as a function of time, inferred from the light curve of Figure~\ref{fig:lightcurve} and estimates of the normal brightness of the four stars from stellar evolution models (see the text). The bottom panel displays the same extinction curve as a function of the radial distance of star Aa from the center, for comparison with the boundaries of the gas and dust disks as reported by \cite{Kennedy:2019}. The hatched areas indicate the formal uncertainty in the location of the edges of the gas disk. Those for the dust disk are negligible on the scale of this figure. \label{fig:extinction}}
\end{figure}

The result of this exercise is shown in the top panel of Figure~\ref{fig:extinction}.  Binary A became too faint to influence the ASAS-SN observations around 2026\,May\,$6 \pm 2$, once the extinction reached 4 or 5\,mag.  We are not able to estimate further increases in the extinction beyond that point, as the light contribution from Aa+Ab became negligible compared to the measurement errors. The bottom panel of Figure~\ref{fig:extinction} displays the extinction as a function of the distance of star Aa from the center of mass of the B binary (i.e., the assumed center of the disk). For simplicity, here we have neglected the slightly non-circular sky-projected shapes of the gas and dust components of the disk. 

Two interesting facts from the bottom panel of this figure are worth mentioning. The first is that significant blocking of the light from star Aa began as it was already well behind the gaseous component of the disk, rather than at its outer edge, or the outer edge of the dust component. This is not unexpected, as mentioned earlier, because the gas can trap small particles capable of scattering and absorbing light at optical or near-infrared wavelengths.
Secondly, the plot shows that the bulk of the extinction over the range we can measure it occurred over period of about three weeks, and a travel distance of only 0.4\,au, implying a rather sharp boundary for the absorbing material at approximately 5.5\,au. It is not known whether extinction continues to rise inward, or whether it fluctuates depending on the local density of the material, due to ripples, warps, or spiral arms in the disk.
Fine structure in circumbinary disks has been observed before, such as in the well-known PMS binary KH\,15D \citep[see, e.g.,][and references therein]{Lamitina:2025, Hamilton:2026}, and would not be surprising to see in a more complicated quadruple system such as \hd.

We note, in passing, that the extinction at the time of our second CHARA image, in which component Ab became invisible, was only about 0.5\,mag in $g$ (Figure~\ref{fig:extinction}, top panel) and even less at near infrared wavelengths. This is not enough to explain the total disappearance of the star. However, as shown in Figure~\ref{fig:approach}, and also in the next section, at the time of the CHARA observation component Ab was well ahead of star Aa on its path inward, by roughly 1\,au. It is therefore likely to have experienced much greater attenuation than component Aa. {Indeed, as seen in that figure, star Ab was already inside of the $\sim$5.5\,au boundary at which star Aa later disappeared (inner dashed line), whereas star Aa was still outside.}

\subsection{Possible eclipse of star Ab}
\label{sec:eclipseAb}

As seen in Figure~\ref{fig:approach}, the projected path of star~Ab relative to the center of mass of \hdb\ brings it very close to the projection of the orbit of component Ba on the plane of the sky. As unlikely as it may seem a priori, this suggests the possibility of an eclipse of Ab, provided the two stars are near their locations of closest approach of their projected paths at the same time.

\begin{figure}
\includegraphics[width=\columnwidth]{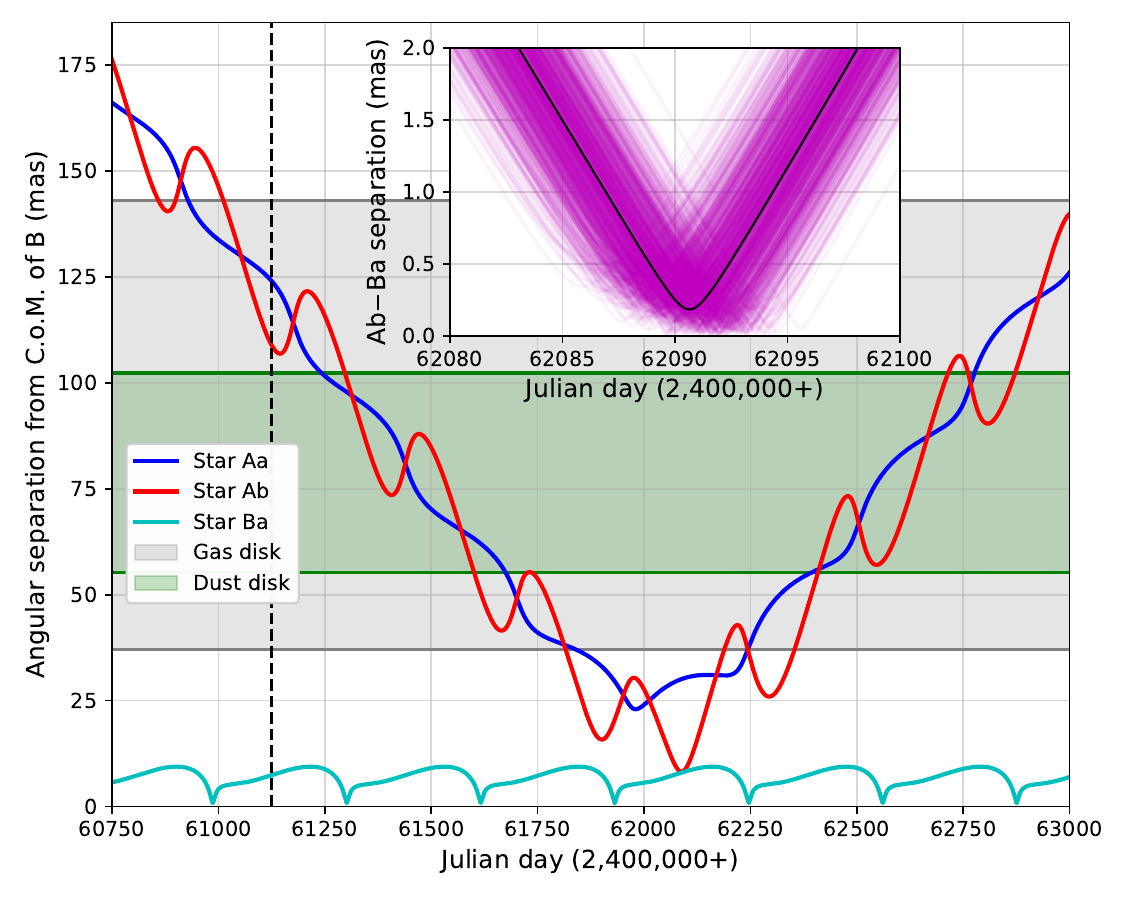}
\figcaption{Projected angular separation of stars Aa, Ab, and Ba from the center of mass of \hdb, based on the present orbital analysis. 
 Also indicated are the boundaries of the dust and gas components of the disk, taken from the results of \cite{Kennedy:2019}. The vertical dashed line marks the date of our 2026\,Mar\,25 CHARA image, when star Ab was closer to the center of the disk than Aa. The Ab and Ba curves come close to contact at a time near the middle of the range shown, indicating that an eclipse of Ab is possible. The inset shows the projected separation between stars Ab and Ba as a function of time, which reaches a minimum of only $\sim$0.2\,mas on JD~2,462,091 (2028~November~15). The shaded magenta area represents the uncertainty in the separation predictions (see the text). \label{fig:eclipseAb}}
\end{figure}

In Figure~\ref{fig:eclipseAb}, we plot the time dependence of the angular separations of stars Ab and Ba from the center of mass of the B binary, and indeed their predicted paths based on the present orbital solutions come within $\sim$0.2 mas of each other on 2028~November~15 (JD~2,462,091), with an uncertainty of two or three days. To illustrate the uncertainty in their actual sky-projected separation, the inset shows 1000 realizations calculated from sets of orbital elements taken directly from our posterior distributions. While the likelihood of a central eclipse appears small, we cannot currently rule out the possibility of a partial eclipse, especially considering that the angular sizes of the stars (with diameters of about 0.21 and 0.16\,mas for Ba and Ab) are comparable to the minimum separation. The two stars will be moving in the same direction at the time of the encounter (essentially due north), which would contribute to lengthening the duration of the event. However, as star Ab is the fainter of the four components, the flux decrement for the combined light of the quadruple system would only be slight ($\sim$0.01\,mag in $V$, and $\sim$0.08\,mag in $K$, at most).

We conclude this section by pointing out, in reference to Figure~\ref{fig:eclipseAb}, that whereas the inward motion of star Aa is monotonic, the projected distance of Ab from the center of the disk oscillates as the star orbits its primary. As a result, star Ab can suffer more or less extinction than the primary at certain times. On the date of our 2026\,Mar\,25 CHARA image (dashed line), star Ab was closer to the center and behind more of the absorbing material, as mentioned before. Three to four months later the opposite was true, and indeed, evidence from speckle observations in mid July (A.\ Tokovinin 2026, private communication) indicates that star Ab was temporarily visible again while Aa was further in, and either absent or very much fainter.

\section{Final Remarks}
\label{sec:conclusions}

In the nearly four decades since it was first recognized as an object worthy of attention, the quadruple PMS system \hd\ has been the subject of numerous investigations ---both observational and theoretical--- that have gradually improved  our understanding of its properties. Beginning in 2026, it is rewarding astronomers with a rare display that had not been anticipated a decade ago, which is the passage of the A binary directly behind the circumbinary disk of \hdb. This is a result of the fortuitous orientation of the $\sim$240\,yr outer orbit, which is within just 2\arcdeg\ of being edge-on. Studies of the structure and composition of the disk are already being undertaken, and as the event progresses, the interpretation of the observations will depend to some extent on an accurate knowledge of the dynamics of the system, i.e., the three orbits and the masses of the four stars.

To benefit current and future studies of this unique event, here we have reported numerous new spectroscopic and astrometric observations that we have incorporated into an updated global orbital analysis of the system. They include: (i) extensive RV monitoring of the three visible stars, now covering a time span of more than 36\,yr; (ii) long-baseline interferometry of both \hda\ and B from the KI as well as the CHARA Array and the VLTI, now spanning a total of 23\,yr; and (iii) recent astrometric measurements of the outer orbit from the CHARA Array, the VLTI, and additional speckle interferometry observations. We have substantially improved most of the inner and outer orbital elements, and have determined the masses with uncertainties that range from 1.2 to 2.6\% for the four components. The distance to the system is now known to 0.8\%. These properties, together with effective temperature determinations from the literature, and our brightness estimates, are shown to agree well with current models of stellar evolution at the current 10\,Myr age of the parent TW\,Hya association.

We present $N$-body simulations to quantify changes in the orbital elements of A and B as they encountered each other during the recent 2021 periastron passage in the outer orbit, and also changes due to the influence of the circumbinary disk. We estimate that those changes are below the level at which they can make a difference in our results, given the current observational uncertainties. {Over the long term up to the age of 10\,Myr, we find that the system is dynamically stable, but not static.}

A near-infrared image we obtained using the CHARA Array, as well as ground-based optical photometry from ASAS-SN, show the eclipse of the A binary to be underway.  Binary A is effectively invisible in the optical at the time of writing, resulting in a drop in the total brightness of \hd\ of a full magnitude. We use this information to infer changes in the optical extinction from the disk, and to show that the boundary at which most of the extinction sets in is located at about 5.5\,au from the center,
at least on the ingress side. The boundary appears to be quite sharp, as binary A went from its normal brightness to near complete disappearance in about three weeks' time, over a travel distance of $\sim$0.4\,au. Finally, we point out that based on the current orbits the sky-projected paths of stars Ab and Ba will come tantalizingly close to each other in mid November of 2028 (within about 0.2\,mas), and we comment briefly on the slight possibility of at least a partial eclipse of star Ab.


\acknowledgements

We are grateful to P.\ Berlind, M.\ Calkins, J.\ Caruso, R.\ Davis,
G.\ Esquerdo, E.\ Horine, A.\ Milone, J.\ Peters, and J.\ Zajac for
their help in obtaining the spectroscopic observations of \hd\ reported here.
Some of those observations were obtained at the Smithsonian
Astrophysical Observatory's Fred Lawrence Whipple Observatory Ridge
Telescopes. We also thank A.\ Tokovinin (CTIO) and S.\ Andrews (SAO) for
helpful discussions, {and the anonymous referee for constructive comments.}
This work is based in part upon observations obtained with the Georgia State University Center for High Angular Resolution Astronomy Array at Mount Wilson Observatory.  The CHARA Array is supported by the National Science Foundation under Grant No. AST-2034336 and AST-2407956. Institutional support has been provided from the GSU College of Arts and Sciences, Office of the Provost, and Office of the Vice President for Research and Economic Development.
{SK acknowledges funding for MIRC-X from the European Research Council (ERC) under the European Union's Horizon 2020 research and innovation programme (Starting Grant No.\ 639889 and Consolidated Grant No.\ 101003096). JDM acknowledges funding for the development of MIRC-X (NASA-XRP NNX16AD43G, NSF-AST 2009489) and MYSTIC (NSF-ATI 1506540, NSF-AST 1909165, NSF-AST 2406873).}
{EH acknowledges support from NSF grant AST-2510385.}
{This research has made use of the Jean-Marie Mariotti Center \texttt{Aspro} service (available at \url{http://www.jmmc.fr/aspro}), the \texttt{SearchCal} service (\url{https://www.jmmc.fr/searchcal}, and the OiDB service (\url{http://oidb.jmmc.fr}).} {The work is based in part on observations collected at the European Southern Observatory under ESO programs 113.26ML.001 and 113.26ML.006.}
The research has made extensive use of the SIMBAD and VizieR
databases, operated at the CDS, Strasbourg, France, and of NASA's
Astrophysics Data System Abstract Service. We also acknowledge the use
of the Washington Double Star Catalog, maintained at the U.S. Naval
Observatory.  R.\ Matson (USNO) kindly provided the
individual astrometric measurements of \hd\ from the WDS. 
The computational resources used for this research include the Smithsonian
High Performance Cluster (SI/HPC), Smithsonian Institution
(\url{https://doi.org/10.25572/SIHPC}).

\section*{Appendix~A: Component Magnitude Differences Within \hda\ and B}
\label{sec:appendixA}

\begin{figure}
  \includegraphics[width=0.5\textwidth]{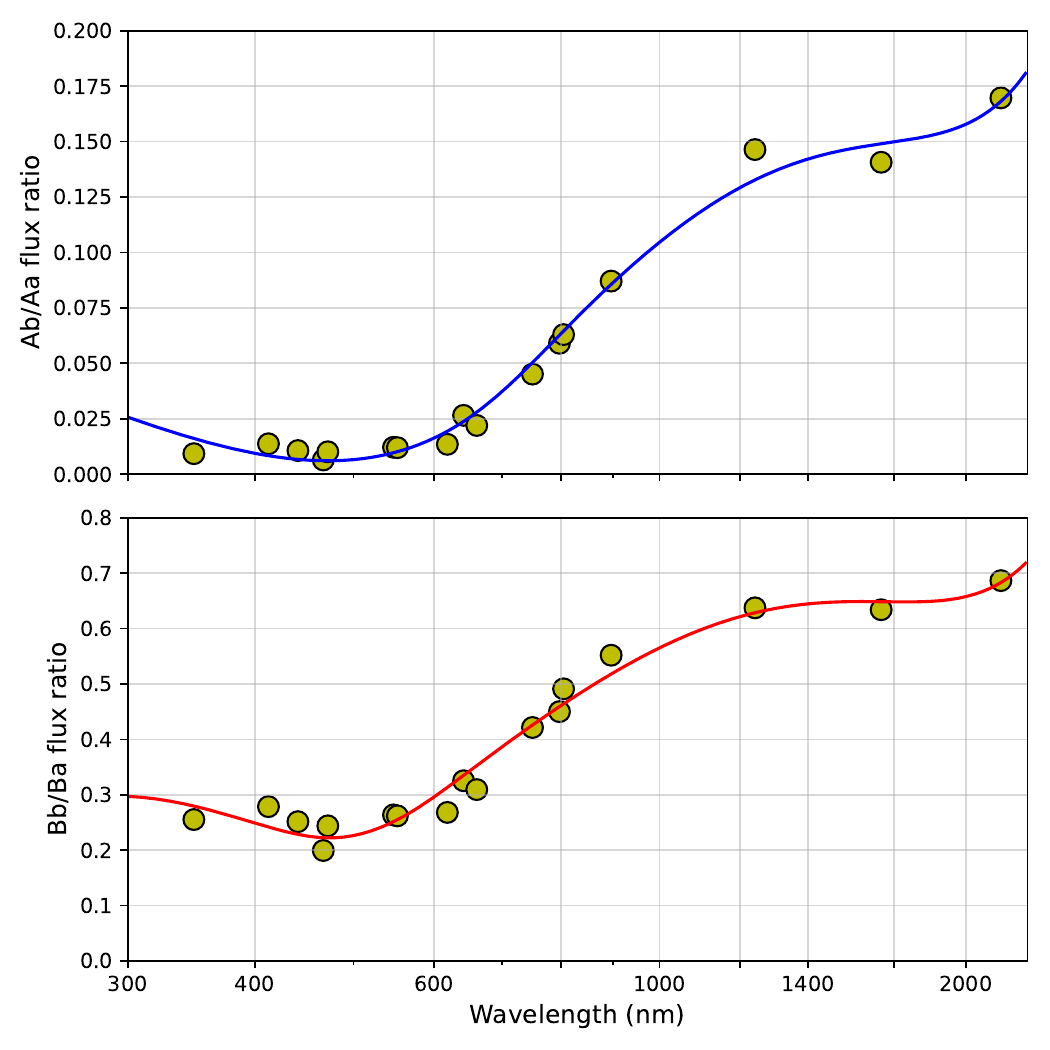}
  \figcaption{Calculated flux ratios between the primary and secondary within each subsystem of \hd. The points represent values obtained from synthetic spectra in 17 different photometric filters. The curves are cubic spline fits to those data, to a level of approximation sufficient for our purposes.
    \label{fig:fluxratioAB}}
\end{figure}

{The brightness differences between the primary and secondary within \hda\ and \hdb\ as a function of wavelength were calculated with the aid of synthetic spectra from the work of \cite{Husser:2013} based on PHOENIX models. For binary A, we adopted temperatures of 4500 and 3000\,K, surface gravities of $\log g = 4.0$ for both stars, and solar metallicity. For binary B, the temperatures were 4200 and 3600\,K. Synthetic flux ratios were calculated for 17 representative passbands in standard photometric systems (Str\"omgren, SDSS, Johnson-Cousins, TESS, Kepler, and 2MASS), and the trends were approximated with cubic spline functions, shown in Figure~\ref{fig:fluxratioAB}. These were then used to infer the magnitude differences $\Delta m$ required in eq.[\ref{eq:photocenter}] at the wavelength of each visual or speckle observation.
We note as a consistency check that the theoretical values for the flux ratios in the $H$ band are in good agreement with those determined empirically from the interferometric measurements in this work and in that of \cite{Zuniga:2021, Zuniga:2026}, which are $0.14 \pm 0.02$ and $0.65 \pm 0.03$ for \hda\ and \hdb\ (Section~\ref{sec:stellarevolution}).}

\section*{Appendix~B: KI Visibility Measurements}
\label{sec:appendixB}

We present here the figures showing the recalibrated KI visibility measurements, compared to our model predictions for \hda\ and B.

\begin{figure*}
\includegraphics[width=\textwidth]{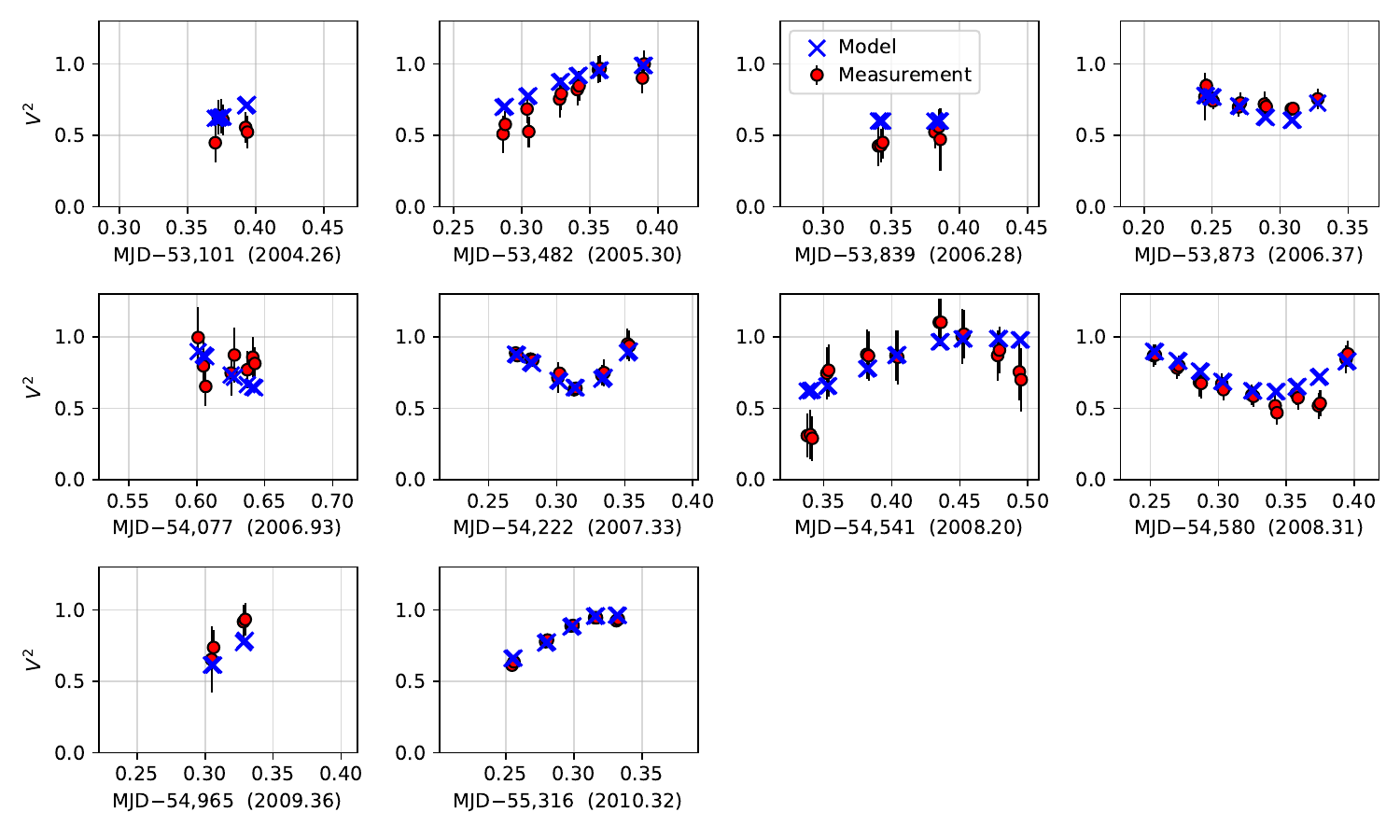}
\figcaption{Visibility measurements for the 10 nights of KI observations of \hda, compared with our model predictions. The measurement uncertainties shown include a jitter term added in quadrature to the internal errors, to account for atmospheric effects or other instrumental systematic errors not corrected by the calibration pipeline.\label{fig:V2A}}
\end{figure*}

\begin{figure*}
\includegraphics[width=\textwidth]{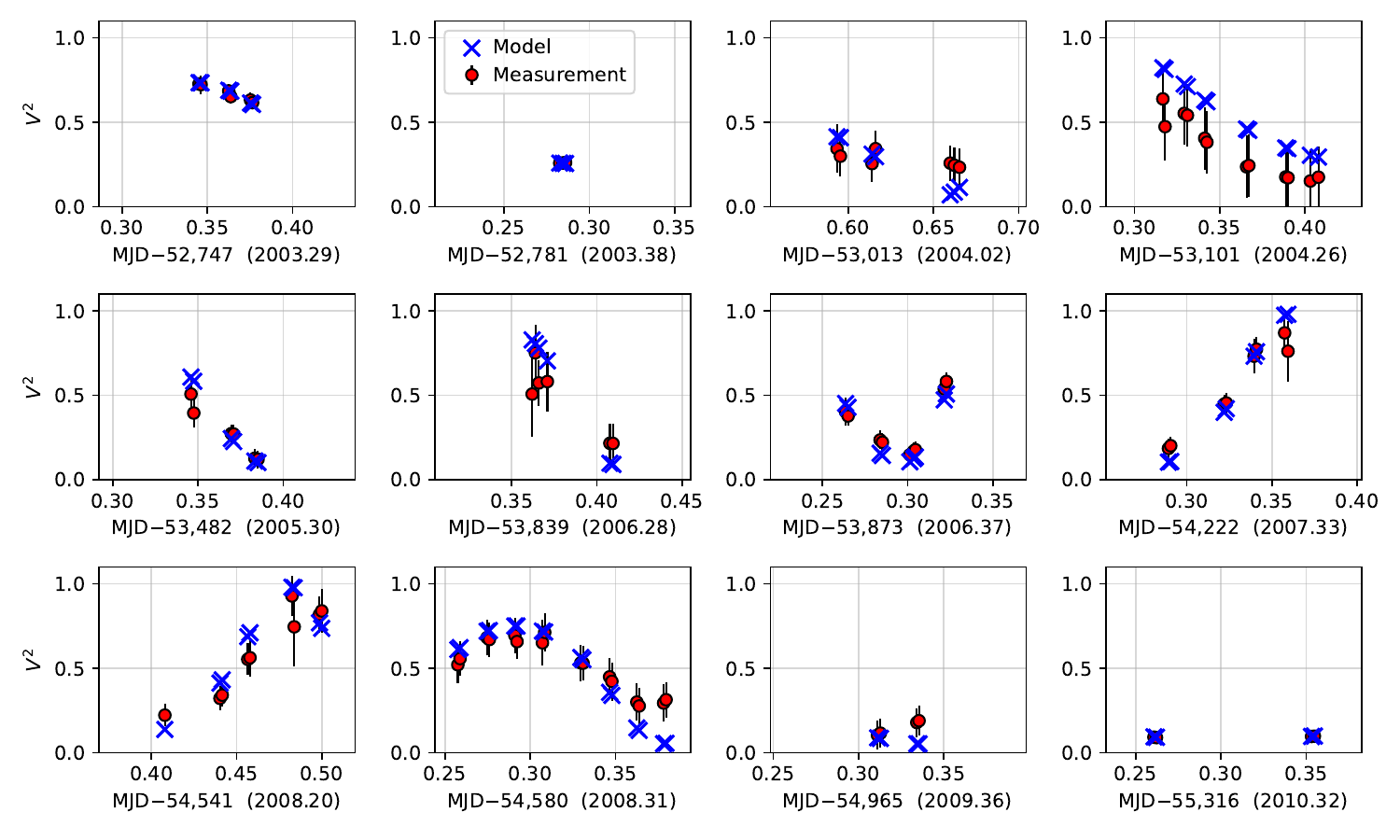}
\figcaption{Similar to Figure~\ref{fig:V2A}, for the 12 nights of KI observations of \hdb. \label{fig:V2B}}
\end{figure*}

\clearpage


\end{document}